\documentclass[aps,longbibliography,nofootinbib]{revtex4-1}
\usepackage[latin1]{inputenc} 
\usepackage{amssymb}
\usepackage{amsfonts} 
\usepackage{amsthm} 
\usepackage{amsmath} 
\usepackage{epsfig}
\usepackage{hyperref}
\usepackage{color}
\usepackage{bbold}
\usepackage[T1]{fontenc}
\usepackage{young}
\usepackage{youngtab}
\usepackage{braket}

\def\nn{\nonumber}
\def\ic{\mathrm{i}}

\def \bc {\begin{center}}
\def \ec {\end{center}}
\def \bi {\begin{itemize}}
\def \ei {\end{itemize}}

\def \ba {\begin{array}}
\def \ea {\end{array}}

\def \bea {\begin{eqnarray}}
\def \eea {\end{eqnarray}}

\def \be {\begin{equation}}
\def \ee {\end{equation}}

\newcommand{\la}{\langle}
\newcommand{\ra}{\rangle}

\def \um {\frac{1}{2}}

\def\tr {\mathrm{tr}}

\def\bc {\bar{\beta}}

\def\nb{{\vec{n}}}
\def\zb{\mathbf{z}}
\def\eb{\mathbf{e}}
\def\E{E}
\def\rmu{\mathrm{U}}
\def\dcat{{\scriptstyle\mathrm{DCAT}}}
\def\2cat{{\scriptstyle\mathrm{2CAT}}}
\def\3cat{{\scriptstyle\mathrm{3CAT}}}
\def\noon{{\scriptstyle\mathrm{NOON}}}
\def\nodon{{\scriptstyle\mathrm{NODON}}}

\begin{document}

\title{Entanglement and U(D)-spin squeezing in symmetric multi-quDit systems and applications to quantum phase transitions 
in Lipkin-Meshkov-Glick D-level atom models}

\author{Manuel Calixto}
\email{calixto@ugr.es}
\affiliation{Department of Applied Mathematics and Institute Carlos I of Theoretical and Computational Physics (iC1), University of  Granada,
Fuentenueva s/n, 18071 Granada, Spain}
\author{Alberto Mayorgas}
\email{albmayrey97@gmail.com}
\affiliation{Department of Applied Mathematics and Institute Carlos I of Theoretical and Computational Physics (iC1), University of  Granada,
Fuentenueva s/n, 18071 Granada, Spain}
\author{Julio Guerrero}
\email{jguerrer@ujaen.es}
\affiliation{Department of Mathematics, University of Jaen, Campus Las Lagunillas s/n, 23071 Jaen, Spain}

\date{\today}

\begin{abstract}
Collective spin operators for symmetric multi-quDit (namely, identical $D$-level atom) systems generate a U$(D)$ symmetry. We explore generalizations  to arbitrary $D$ of SU(2)-spin coherent states and 
their adaptation to parity (multicomponent Schr\"odinger cats), together with multi-mode extensions of NOON states. We write level, one- and two-quDit reduced density matrices of symmetric $N$-quDit states, 
expressed in the last two cases in terms of collective U$(D)$-spin operator expectation values. 
Then we evaluate level and particle entanglement for symmetric multi-quDit states with linear and von Neumann entropies of the corresponding reduced density matrices. In particular, we analyze the numerical and variational ground state of 
Lipkin-Meshkov-Glick  models of $3$-level identical atoms.  We also propose an extension of the concept of 
SU(2) spin squeezing to SU$(D)$ and relate it to pairwise $D$-level atom entanglement. Squeezing parameters and entanglement entropies  are good markers that characterize the different quantum phases, 
and their corresponding critical points, that take place in these interacting $D$-level atom models.
\end{abstract}

%
%
%
%
%


\maketitle

\section{Introduction}

The development of quantum technologies partially relies on the efficient preparation of nonclassical atomic states and the exploitation of many-body entanglement  
\cite{RevModPhys.80.517-Amico,RevModPhys.90.035005-Pezze,RevModPhys.81.865-Horodecki} and spin squeezing \cite{MA201189}, specially to enhance the sensitivity of precision measurements like in quantum metrology. 
Such is the case of many-body entangled (and spin-squeezed) states of cold atoms generated for instance in atom-atom collisions in Bose-Einstein
condensates (BECs) \cite{RevModPhys.90.035005-Pezze}. 

Indistinguishable particles are naturally correlated due to exchange symmetry and there has been a long-standing debate on whether identical particle entanglement is physical or merely a
mathematical artifact (see e.g \cite{BENATTI20201} and references therein). Recent work like  \cite{PhysRevX.10.041012} shows indeed entanglement between identical particles 
as a consistent quantum resource in some typical optical and cold atomic systems with immediate practical impact. It can also be extracted and used as a resource for standard quantum
information tasks \cite{PhysRevLett.112.150501_Plenio}. Moreover, multipartite entanglement 
of symmetric multi-qubit systems can add robustness and stability against the loss of a small number of particles \cite{BENATTI20201}.

Understanding the role of the indistinguishableness of identical bosons and quantum entanglement has been the subject of many recent work 
(see e.g. \cite{Dalton_2017,Dalton_2017_2} and references therein). We know that, for $N=2$ particles, any quantum state is
either separable or entangled. However, for $N > 2$, one needs further classifications for multipartite entanglement \cite{PhysRevA.62.062314-DurVidalCirac}. 
Many different measurements have been proposed to detect and quantify quantum correlations \cite{RevModPhys.81.865-Horodecki}. We shall restrict ourselves to bipartite 
entanglement of pure states, where necessary conditions for separability  in arbitrary dimensions exist. 

In order to quantify entanglement between identical particles we shall follow  Wang and M\o{}lmer's \cite{Molmer} procedure, who wrote the
reduced density matrix (RDM) of one- and two- qubits, extracted at random from  a symmetric multi-qubit state $\psi$,  in terms of expectation values $\la \vec{S}\ra_\psi$ of collective spin
operators $\vec{S}$. For pairwise entanglement, the concurrence $C$ (an entanglement measure introduced by Wootters \cite{PhysRevLett.80.2245}) was calculated for spin
coherent states (SCSs) \cite{Radcliffe,GilmorePhysRevA.6.2211},  Dicke and  Kitagawa-Ueda \cite{Kitagawa} spin squeezed states, together with mixed states of  Heisenberg models.  
Kitagawa-Ueda \cite{Kitagawa} spin squeezed states are the spin version of  traditional parity adapted CSs, sometimes called ``Schr\"odinger cat states'' since they are a quantum superposition of weakly-overlapping 
(macroscopically distinguishable) quasi-classical coherent wave packets. They where first introduced by Dodonov, Malkin and Man'ko  \cite{Dodonovcat} and later adapted to more general finite groups than the parity 
group $\mathbb{Z}_2=\{1,-1\}$ \cite{Mankocat}. In this article we shall introduce $\rmu(D)$ SCSs (denoted DSCSs for brevity) adapted to the parity symmetry $\mathbb{Z}_2\times\stackrel{D-1}{\dots}\times\mathbb{Z}_2$, 
which are a $D$-dimensional generalization of $\rmu(2)$ Schr\"odinger cats, and we shall refer to them as $\dcat$s for short. 
In general, parity adapted CSs are a special set of ``nonclassical'' states with interesting statistical properties (see \cite{M.Nieto&D.Traux,V.Buzek&A.Viiella-Barranco&P.Knight,M.Hillery} for several seminal papers). 
Parity adapted DSCSs arise as variational states reproducing the energy and structure of ground states in  Lipkin-Meshkov-Glick (LMG) $D$-level atom models (see \cite{nuestroPRE} and later in Sec. \ref{LMGsec}). 

In Ref. \cite{PhysRevA.68.012101-Wang}, the concurrence $C$ was related to the spin squeezing parameter $\xi^2=4(\Delta S_{\vec{n}_\perp})^2/N$ introduced by \cite{Kitagawa}, which measures 
spin fluctuations in an orthogonal direction to  the mean value $\vec{n}\propto \la \vec{S}\ra$ with minimal variance. Spin squeezing means that $\xi^2<1$, that is, when  the variance $(\Delta S_{\vec{n}_\perp})^2$ 
is smaller than the standard quantum limit $S/2=N/4$ (with $S$ the spin) attained by (quasiclassical) SCSs. This study shows that spin squeezing is related to
pairwise correlation for even and odd parity multi-qubit states. Squeezing is in general a redistribution of quantum fluctuations
between two noncommuting observables $A$ and $B$ while preserving
the minimum uncertainty product $\Delta_\psi A\Delta_\psi B\geq \um |\la [A,B]\ra_\psi|$. Roughly speaking, it means to partly cancel out
fluctuations in one direction at the expense of those enhanced in the ``conjugated'' direction. For the standard radiation field, it implies the variance relation $(\Delta q)^2<1/4$ for 
quadrature (position $q$ and momentum $p$) operators. For general $\rmu(D)$ spin systems of identical $D$-level atoms or ``quDits'', the situation is more complicated and we shall extend the $D=2$ definition of spin squeezing to 
general $D$.

Spin squeezing can be created in atom systems by making them to interact with each other for a
relatively short time in Kerr-like medium with ``twisting'' nonlinear Hamiltonians like $H=\lambda S_x^2$  \cite{Kitagawa},  
generating entanglement between them \cite{Sorensen2001,PhysRevLett.86.4431-Sorensen}. This effective Hamiltonian can be realized in ion traps \cite{PhysRevLett.82.1835Molmer} 
and has produced four-particle entangled states \cite{Sackett2000}. There are also some proposals for two-component BECs \cite{Sorensen2001}.
Likewise, the ground state at zero temperature of Hamiltonian critical many-body systems possessing discrete (parity) symmetries also exhibits a cat-like structure. The parity symmetry is spontaneously 
broken in the thermodynamic limit $N\to\infty$ and degenerated ground states arise. Parity adapted coherent states are then good variational states, reproducing the energy of the ground state 
of these quantum critical models in the thermodynamic limit $N\to\infty$, namely in matter-field interactions (Dicke model) of two-level \cite{PhysRevA.84.013819,PhysRevA.85.053831} and 
three-level \cite{PhysRevA.92.053843,L_pez_Pe_a_2015} atoms, BEC \cite{citlali}, $\rmu(3)$ vibron models of molecules \cite{Calixto_2012,PhysRevA.89.032126}, bilayer quantum Hall systems \cite{Calixto_2018} and  (LMG) models for 
two-level atoms \cite{octavio,Romera_2014,Calixto_2017}. Quantum information (fidelity, entropy, fluctuation, entanglement, etc) measures  have proved to be useful in the analysis 
of the highly correlated ground state structure of these many-body systems and the identification of critical points across the phase diagram. Special attention must be paid to the deep  
connection between entanglement, squeezing and quantum phase transitions (QPTs); see \cite{RevModPhys.80.517-Amico,MA201189} and references therein.  

In this article we want to explore squeezing and  interparticle and interlevel quantum correlations in symmetric multi-quDit systems like the ones described by critical LMG models of identical $D$-level atoms (se e.g. 
\cite{Kus,KusLipkin,Meredith,Casati,Saraceno,Meredith,nuestroPRE} for $D=3$ level atom models). The literature mainly concentrates on two-level atoms, displaying a $\rmu(2)$ symmetry, which is justified 
when we make atoms to interact with an external monochromatic electromagnetic field. However, the possibility of polychromatic radiation requires the activation of more atom levels and increases the 
complexity and richness of the system (see e.g. \cite{PhysRevA.92.053843}). In any case, this also applies to general interacting boson models \cite{iachello_arima_1987} and 
multi-mode BECs with two or more boson species. Collective operators generate a $\rmu(D)$ spin symmetry for the case of $D$-level identical atoms or $D$ boson 
species (quDits). Recently \cite{nuestroPRE} we have calculated the phase diagram of a  three-level LMG  atom model. Here we want to explore the connection between entanglement and squeezing with QPTs for this symmetric multi-qutrit system. 
For this purpose, we extend to $D$ levels the usual definition of Dicke, parity-adapted SCSs and $\noon$ states, and propose linear and von Neumann entropies of certain reduced density matrices as a measure of interlevel and interparticle entanglement. We also introduce a generalization of $\mathrm{SU}(2)$ spin squeezing to $\mathrm{SU}(D)$.

The organization of the paper is as follows. In Sec. \ref{statesymmatsec} we introduce collective $\rmu(D)$ spin operators $S_{ij}$, their boson realization, 
their matrix elements in Fock subspaces of $N$ symmetric quDits, DSCSs and their adaptation to  parity (``$\dcat$s''), and a generalization of $\noon$ states to $D$-level systems (``$\nodon$s''). 
In Sec. \ref{entangsec} 
we give a brief overview on the concepts and measures
of interlevel and interparticle entanglement, considering different bipartitions of the whole system,  that we put in practice later in Sec. \ref{signaturesec}.  
As entanglement measures, we concentrate on linear (unpurity) and von Neumann entropies. We compute entanglement between levels and atoms for DSCSs, $\dcat$ and $\nodon$ states. 
In Sec. \ref{squeezingsec} we extend Kitagawa-Ueda's definition of $\mathrm{SU}(2)$ spin squeezing to  $\mathrm{SU}(D)$ and we also connect it with two-quDit entanglement introduced in the previous Section.  
 In Sec. \ref{LMGsec} we introduce $D$-level Lipkin-Meshkov-Glick (LMG) atom models (we particularize to $D=3$ for simplicity) and study their phase diagram and critical points. In Sec. \ref{signaturesec} we 
analyze the ground state structure of the three-level LMG atom model across the phase diagram  with the quantum information measures of Sec. \ref{entangsec} and 
the $\mathrm{SU}(3)$-spin squeezing parameters of Sec. \ref{squeezingsec}, 
thus revealing the role of entanglement and squeezing as  signatures of quantum phase transitions and  detectors of critical points. We only compute linear entropy in Sec. \ref{signaturesec}, 
since it is easier to compute than von Neuman entropy  and eventually provides similar qualitative information for our purposes; the interested reader can consult Refs. 
\cite{JPhysA.36.12255,PhysRevA.67.022110} for a more general study on the relation between both entropies. Finally, Sec. \ref{conclusec} is devoted to conclusions.

\section{State space, symmetries and collective operator matrix elements}\label{statesymmatsec}

 We consider a system of $N$ identical atoms of $D$ levels ($N$ quDits in the quantum information jargon). Let us denote by 
 $E_{ij}=|i\rangle\langle j|$ the Hubbard operator describing a 
transition from the single-atom level $|j\rangle$ to the level $|i\rangle$, with $i,j=1,\dots,D$.  These are a generalization of Pauli matrices for qubits ($D=2$), namely  
$\E_{12}=\sigma_+, \E_{21}=\sigma_-, \E_{11}-\E_{22}=\sigma_3$ and 
$\E_{11}+\E_{22}=\sigma_0$ (the $2\times 2$ identity  matrix). 
The expectation values of $E_{ij}$ account for  complex polarizations or coherences between levels for $i\not=j$ and 
occupation probability of the level $i$ for $i=j$. The  $E_{ij}$ represent the $D^2$ step operators 
of $\rmu(D)$, whose (Cartan-Weyl) matrices 
$\langle l|E_{{ij}}|k\rangle=\delta _{{il}}\delta _{{jk}}$ fulfill the  commutation relations 
\be
\left[E_{{ij}},E_{{kl}}\right]=\delta _{{jk}} E_{{il}} -\delta _{{il}} E_{{kj}}.\label{commurel}
\ee
Let us denote  by $E_{ij}^\mu$, $\mu=1,\dots,N$ the embedding of the single $\mu$-th atom  $E_{ij}$ operator into the $N$-atom Hilbert space; namely, 
$E_{ij}^3=\mathbb{1}_D\otimes \mathbb{1}_D\otimes E_{ij}\otimes \mathbb{1}_D$ for $N=4$, with $\mathbb{1}_D$ the $D\times D$ identity matrix. The collective $\rmu(D)$-spin  ($D$-spin for short) 
operators are 
\be 
S_{ij}=\sum_{\mu=1}^N E_{ij}^\mu, \quad i,j=1,\dots,D.\label{collectiveS}
\ee
They are the generators of the underlying $\rmu(D)$ dynamical symmetry with the same commutation relations 
as those of $E_{ij}$ in \eqref{commurel}. When focusing on  two levels $i>j$, we might  prefer to use 
\be
\vec{J}^{(ij)}=(J^{(ij)}_x,J^{(ij)}_y,J^{(ij)}_z), \quad J^{(ij)}_x=\frac{S_{ij}+S_{ji}}{2},\quad 
J^{(ij)}_y=\ic \frac{S_{ij}-S_{ji}}{2}, \quad J^{(ij)}_z=\frac{S_{jj}-S_{ii}}{2},\label{su2embedding}
\ee
(roman $\ic$ denotes the imaginary unit throughout the article) with commutation relations $[J^{(ij)}_x,J^{(ij)}_y]=\ic J^{(ij)}_z$ (and cyclic permutations of $x,y,z$), which is an embedding of $D(D-1)/2$ $\mathrm{SU}(2)$ subalgebras into $\rmu(D)$. 
Although the form \eqref{su2embedding} for $D$-spin operators could be more convenient to extrapolate all the $D=2$ level machinery to arbitrary $D$, we shall still prefer the form \eqref{collectiveS} (at least 
in this paper), which allows for more compact formulas.

The  $D^N$-dimensional Hilbert space is the $N$-fold tensor product $[\mathbb{C}^D]^{\otimes N}$. 
The tensor product representation of $\rmu(D)$ is reducible and decomposes into a Clebsch-Gordan direct sum of mixed symmetry invariant subspaces. Here we shall restrict ourselves to the $\tbinom{N+D-1}{N}$-dimensional 
fully symmetric sector (see \cite{nuestroPRE} for the role of other mixed symmetry sectors), which means that our $N$ atoms are indistinguishable (bosons). Denoting by $a^\dag_i$ (resp. $a_i$) the 
creation (resp. annihilation) operator of an atom in the $i$-th level, the collective $D$-spin operators \eqref{collectiveS} can be expressed (in this fully symmetric case) as bilinear products of creation and annihilation operators as 
(Schwinger representation)
\be
S_{ij}=a_i^\dag a_j, \quad i,j=1,\dots,D.\label{collop}
\ee
$S_{ii}$ is the operator number of atoms at level $i$, whereas $S_{ij}, i\not=j$ are raising and lowering operators. 
The fully symmetric representation space of $\rmu(D)$ is embedded into Fock space, with Bose-Einstein-Fock basis ($|\vec{0}\ra$ denotes the Fock vacuum)
\be
|\vec{n}\ra=|n_1,\dots, n_D\ra=
\frac{(a_1^\dag)^{n_1}\dots(a_D^\dag)^{n_D}}{\sqrt{n_1!\dots n_D!}}|\vec{0}\ra, \label{symmetricbasis}
\ee
when fixing  $n_1+\dots+n_D=N$ (the linear Casimir $C_1=S_{11}+\dots+S_{DD}$) to the total  number $N$ of atoms. There are other realizations of $D$-spin operators in terms of more than $D$ bosonic modes 
(e.g. when each level $j$ contains $M$ degenerate orbitals), which describe mixed symmetries 
\cite{MOSHINSKY1962384,MOSHINSKY1963173,Biedenharn}, but we shall not consider them here.

Collective $D$-spin operators  \eqref{collop} matrix elements are given by
\be
\la\vec{m}|S_{ij}|\vec{n}\ra=\sqrt{(n_i+1)n_j}\delta_{m_i,n_{i}+1}\delta_{m_j,n_{j}-1}\prod_{k\not=i,j}\delta_{m_k,n_k},\; \forall i\not=j, 
\quad\quad \la\vec{m}|S_{ii}|\vec{n}\ra=n_i\delta_{\vec{m},\vec{n}}.\label{Sijmatrix}
\ee
The expansion of a general symmetric $N$-particle state $\psi$ in the Fock basis will be written as
\be
|\psi\ra=\sum_{\vec{n}}{}'\,c_{\vec{n}}|\vec{n}\ra=\sum_{n_1+\dots+n_D=N} c_{n_1,\dots,n_D}|n_1,\dots, n_D\ra,\label{psisym}
\ee
where $\sum'$ is a shorthand for the restricted sum. $D$-spin operator expectation values (EVs)  can then be easily computed as
\be
\la S_{ij}\ra_\psi=\la\psi| S_{ij}|\psi\ra=\sum_{\vec{n}}{}'\bar{c}_{\vec{n}_{ij}} c_{\vec{n}} \sqrt{(n_i+1)n_j}, \,i\not=j,\quad \la S_{ii}\ra_\psi=\sum_{\vec{n}}{}'n_i |c_{\vec{n}}|^2,\label{isoEV}
\ee
where we have used \eqref{Sijmatrix} and where by $\vec{n}_{ij}$ we mean to replace $n_i\to n_i+1$ and $n_j\to n_j-1$ in $\vec{n}$. 

Among all symmetric multi-quDit states, we shall pay special attention to $\rmu(D)$ SCSs (DSCSs for short)
\be
|\zb\ra=|(z_1,z_2,\dots,z_D)\ra=\frac{1}{\sqrt{N!}}\left(
\frac{z_1a_1^\dag+z_2 a_2^\dag+\dots+z_D a_D^\dag}{\sqrt{|z_1|^2+|z_2|^2+\dots+|z_D|^2}}\right)^{N}|\vec{0}\ra,\label{cohND}
\ee
which are labeled by complex points $\zb=(z_1,\dots,z_D)\in \mathbb{C}^D$. To be more precise, there is an equivalence relation: $|\zb'\ra\sim |\zb\ra$ if $\zb'=q \zb$ for any complex number $q\not=0$, which means that 
$|\zb\ra$ is actually labeled by class representatives of complex lines in $\mathbb{C}^D$, that is, by points of the complex projective phase space 
\[\mathbb CP^{D-1}=[\mathbb{C}^D/\sim]=\rmu(D)/[\rmu(1)\times \rmu(D-1)].\] 
A class/coset representative can be chosen as $\tilde{\zb}= \zb/z_i$ when $z_i\not=0$, which corresponds to a certain patch of the manifold $\mathbb CP^{D-1}$. This is equivalent to chose $i$ as a reference level and set $z_i=1$. 
For the moment, we shall allow redundancy in $\zb$ to write general expressions, although we shall eventually take $i=1$ as a reference (lower energy) level and set $z_1=1$ in Section  \ref{LMGsec}.

DSCSs (multinomial) can be seen as BECs of $D$ modes, generalizing the spin $\rmu(2)$ (binomial) coherent states of two modes introduced by \cite{Radcliffe} and \cite{GilmorePhysRevA.6.2211} long ago.  
For $\zb=\eb_i$ (the standard/canonical basis vectors of $\mathbb{C}^D$), the DSCS $|\eb_i\ra=(a_i^\dag)^{N}|\vec{0}\ra/\sqrt{N!}$ corresponds to a BEC of $N$ atoms placed at level $i$.\footnote{Note the difference 
between Fock states $|n_1,\dots, n_D\ra$ and DSCSs $|(z_1,\dots, z_D)\ra$, which are placed inside parentheses to avoid confusion. For instance, $|\eb_i\ra=|(0,\dots,1,\dots,0)\ra=(a_i^\dag)^{N}|\vec{0}\ra/\sqrt{N!}=|0,\dots,N,\dots,0\ra$.} If we order levels 
$i=1,\dots,D$ from lower to higher energies, the state  $|\eb_1\ra$ would be the ground state, whereas general $|\zb\ra$ could be seen as coherent excitations. Coherent states are sometimes called ``quasi-classical'' states and we shall 
see in Section  \ref{LMGsec} that $|\zb\ra$ turns out to be a good variational state that reproduces the energy and wave function of the ground state of multilevel LMG atom models in the thermodynamic (classical) 
limit $N\to\infty$. 

Expanding the multinomial \eqref{cohND}, we identify the coefficients $c_\nb$ of the expansion \eqref{psisym} of the DSCS $|\zb\ra$ in the Fock basis as
\be
c_\nb(\zb)=\sqrt{\frac{N!}{\prod_{i=1}^D n_i!}}\frac{\prod_{i=1}^D z_i^{n_i}}{|\zb|^{N}},\label{coefCS}
\ee
where we have written $|\zb|=({\zb}\cdot\zb)^{1/2}=(\sum_{i=1}^D |z_i|^2)^{1/2}$ for the length of $\zb$. Note that DSCS are not orthogonal (in general) since 
\be \la \zb'|\zb\ra=\frac{({\zb}'\cdot \zb)^N}{({\zb}'\cdot \zb')^{N/2}({\zb}\cdot \zb)^{N/2}}, \quad {\zb}'\cdot \zb=\bar{z}_1' z_1+\dots+\bar{z}_D' z_D.\label{scprod}
\ee
However, contrary to the standard CSs, they can be orthogonal when ${\zb}'\cdot \zb=0$. EVs of $D$-spin operators $S_{ij}$ (coherences for $i\not=j$ and mean level populations for $i=j$) in a DSCS are simply written as 
\be
\la S_{ij}\ra_\zb=\la \zb|S_{ij}|\zb\ra=N\bar{z}_i z_j/|\zb|^2.\label{CSEV}
\ee
DSCS non-diagonal matrix elements of $D$-spin operators  can also be compactly written as
\be
\la \zb'|S_{ij}|\zb\ra=N \bar{z}'_i z_j\frac{({\zb}'\cdot\zb)^{N-1}}{|\zb'|^N|\zb|^N}.
\ee
Similarly, EVs of quadratic powers of $D$-spin operators in a DSCS state can be concisely written as
\be
\langle \zb|S_{ij}S_{kl}|\zb\rangle=\frac{\bar{z}_iz_l}{|\zb|^4}\left(N\delta_{jk}|\zb|^2+N(N-1)\bar{z}_kz_j\right),\label{qEV}
\ee
and their DSCS matrix elements as
\be
\langle \zb'|S_{ij}S_{kl}|\zb\rangle=\frac{\bar{z}_i'z_l}{|\zb'|^N|\zb|^N}\left(N\delta_{jk}({\zb}'\cdot\zb)^{N-1}+N(N-1)\bar{z}_k'z_j({\zb}'\cdot\zb)^{N-2}\right).
\ee
Note that, for large $N$, quantum fluctuations are negligible and we have $\langle \zb|S_{ij}S_{kl}|\zb\rangle\simeq \langle \zb|S_{ij}|\zb\rangle\langle\zb|S_{kl}|\zb\rangle$. 
Otherwise stated, in the thermodynamical (classical) limit we have
\be
\lim_{N\to\infty}\frac{\langle \zb|S_{ij}S_{kl}|\zb\rangle}{\langle \zb|S_{ij}|\zb\rangle\langle \zb|S_{kl}|\zb\rangle}= 1.\label{nofluct}
\ee
We shall use these ingredients when computing one- and two-quDit RDMs in the next Section. 

We shall see that DSCSs are separable and exhibit no atom entanglement (although they do exhibit level entanglement). The situation 
changes when we deal with parity adapted DSCSs, sometimes called ``Schr\"odinger cat states'' (commented at the introduction), since they are a quantum superposition of weakly-overlapping (macroscopically distinguishable) 
quasi-classical coherent wave packets, as we shall explicitly see below. These kind of cat states arise in several interesting physical situations. As we have already mentioned, they can be generated 
via amplitude dispersion by evolving CSs in  Kerr media, with a strong nonlinear interaction, like the already commented spin-squeezed states of \cite{Kitagawa}. 
They exhibit statistical properties similar to squeezed states, with deviations from Poissonian (CS) distributions. Squeezing and multiparticle entanglement are important 
quantum resources that make Schr\"odinger cats useful for quantum enhanced metrology  \cite{RevModPhys.90.035005-Pezze}. They are also good variational states \cite{GilmorePhysRevA.6.2211}, reproducing the energy of the ground state 
of quantum critical models in the thermodynamic limit $N\to\infty$. To construct them, we require parity operators defined as 
\be \Pi_j=\exp(\ic\pi S_{jj}),\quad j=1,\dots,D.\label{parityop}\ee 
They are conserved when the Hamiltonian scatters pairs of particles conserving the parity of the population $n_j$ in each level $j=1,\dots,D$. 
It is easy to see that $\Pi_j(a^\dag_j)^{n_j}|\vec{0}\ra=(-a^\dag_j)^{n_j}|\vec{0}\ra$, so that the effect of parity operations on number states \eqref{symmetricbasis} is $\Pi_j|\vec{n}\ra=(-1)^{n_j}|\vec{n}\ra$. 
Likewise, using the multinomial expansion  \eqref{cohND}, it is easy to see that the effect of parity operators on symmetric DSCSs  $|\zb\ra$ is then
\be
\Pi_i|\zb\ra=\Pi_i|(z_1,\dots,z_i,\dots,z_D)\ra=|(z_1,\dots,-z_i,\dots,z_D)\ra\label{parityCS}
\ee
Note that $\Pi_i^{-1}=\Pi_i$ and $\Pi_1\dots \Pi_D=(-1)^N$, a constraint that says that the parity group for symmetric quDits is not 
$\mathbb{Z}_2\times\stackrel{D}{\dots}\times\mathbb{Z}_2$ but $\mathbb{Z}_2\times\stackrel{D-1}{\dots}\times\mathbb{Z}_2$ instead. 
In order to define a projector on definite parity (even or odd), we have to chose a reference level, namely $i=1$. Doing so, the projector on even parity becomes
\be
\Pi_\mathrm{even}=2^{1-D}\sum_{\mathbb{b}\in\{0,1\}^{D-1}} \Pi_2^{b_2}\Pi_3^{b_3}\dots \Pi_D^{b_D},\ee
where we denote the binary string $\mathbb{b}=(b_2,\dots,b_D)\in\{0,1\}^{D-1}$. Likewise, the  projection operator on odd parity is $\Pi_\mathrm{odd}=\mathbb{1}-\Pi_\mathrm{even}$. 
Choosing level $i=1$ as a reference level is equivalent to choose a patch on the manifold $\mathbb{C}P^{D-1}$ where $z_1\not=0$; in this way, any 
coherent state $|\zb\ra$ is equivalent to the class representative $|\zb/z_1\ra$, due to  equivalence relation $|\zb'\ra\sim |\zb\ra$ if $\zb'=q \zb$ with $q\neq 0$. Let us simply denote by $\zb=(1,z_2,\dots,z_D)$ 
the class representative in this case. It will be useful, for later use 
as variational ground states, to define the (unnormalized) generalized Schr\"odinger even cat state 
\be
|\dcat\}=\Pi_\mathrm{even}|\zb\ra=2^{1-D}\sum_{\mathbb{b}}|\zb^\mathbb{b}\ra,\label{SC}
\ee
where $\zb^\mathbb{b}=(1,(-1)^{b_2}z_2,\dots,(-1)^{b_D}z_D)$ and we are using $\sum_{\mathbb{b}}$ as a shorthand for  $\sum_{\mathbb{b}\in\{0,1\}^{D-1}}$. It is just the projection of a DSCS on the even parity subspace. 
The state \eqref{SC} is a generalization of the even cat state for $D=2$ in the literature \cite{Dodonovcat}, given by
\be
|\2cat\}=\um \big(|(1,\alpha)\ra+|(1,-\alpha)\ra\big)
\ee
for the class representative $\zb=(z_1,z_2)=(1,\alpha)$. The shorthand $|\alpha\ra=|(1,\alpha)\ra$ is used in the literature when 
a class representative (related to highest $|\eb_1\ra$ or lowest $|\eb_2\ra$ weight fiducial vectors) is implicitly chosen. 
The squared norm of $|\2cat\}$ is simply 
\be
\mathcal{N}(\2cat)^2=\{\2cat|\2cat\}={\um}\left[1+\left(\frac{1-|\alpha|^2}{1+|\alpha|^2}\right)^N\right].
\ee
Note that the overlap $\la 1,\alpha|1,-\alpha\ra=(({1-|\alpha|^2})/({1+|\alpha|^2}))^N\stackrel{N\to\infty}{\longrightarrow} 0$, which means that $|(1,\alpha)\ra$ 
and $|(1,-\alpha)\ra$ are macroscopically distinguishable wave packets for any $\alpha$ (they are orthogonal for $|\alpha|=1$). 
Likewise, the unnormalized $\3cat$ is explicitly given by
\be
|\3cat\}=\frac{1}{4}\big(|(1,\alpha,\beta)\ra+|(1,-\alpha,\beta)\ra+|(1,\alpha,-\beta)\ra+|(1,-\alpha,-\beta)\ra\big)\label{S3C}
\ee
when setting ${\zb}=(z_1,z_2,z_3)=(1,\alpha,\beta)$ as a class representative. The squared norm  is now
\be
\mathcal{N}(\3cat)^2=\frac{1}{4}\left[1+\frac{(1-|\alpha|^2+|\beta|^2)^N+(1+|\alpha|^2-|\beta|^2)^N+(1-|\alpha|^2-|\beta|^2)^N}{(1+|\alpha|^2+|\beta|^2)^N}\right].\label{S3CN}
\ee
These expressions can be generalized to arbitrary $D$ as
\be
\mathcal{N}(\dcat)^2={2^{1-D}}\frac{\sum_{\mathbb{b}}(\zb^\mathbb{b}\cdot\zb)^N}{|\zb|^{2N}}.  
\ee 
We shall use \eqref{S3C} and \eqref{S3CN} in Sections \ref{LMGsec} and \ref{signaturesec}, when discussing a LMG model of atoms with  $D=3$ levels. These $\3cat$ states have also been used 
in $\rmu(3)$ vibron models of molecules \cite{Calixto_2012,PhysRevA.89.032126} and Dicke models of 3-level atoms interacting with a polychromatic radiation field \cite{PhysRevA.92.053843,L_pez_Pe_a_2015}.

The $D$-spin EVs on a $\dcat$ state \eqref{SC} can be now computed and the general expression is
\be
\la \dcat|S_{ij}|\dcat\ra=N\delta_{ij}\frac{\sum_{\mathbb{b}}(-1)^{b_i}|z_i|^2(\zb^\mathbb{b}\cdot\zb)^{N-1}}{\sum_{\mathbb{b}}(\zb^\mathbb{b}\cdot\zb)^N},\label{SDCAT}
\ee
where we set $(-1)^{b_i}= 1=|z_i|$ for $i=1$ (reference level). Similarly, EVs of quadratic powers of $D$-spin operators in a 
$\dcat$ state can be concisely written as
\bea
\la \dcat|S_{ij}S_{kl}|\dcat\ra&=&N(\delta_{ij} \delta_{kl}+\delta_{ik} \delta_{jl}+\delta_{il} \delta_{jk}-2 \delta_{ij}\delta_{jk} \delta_{kl} \delta_{li} )\nonumber\\
&&\times \frac{\sum_{\mathbb{b}}(-1)^{b_i}\bar{z}_iz_l\left[\delta_{jk}
(\zb^\mathbb{b}\cdot\zb)^{N-1}+(N-1)(-1)^{b_k}\bar{z}_kz_j(\zb^\mathbb{b}\cdot\zb)^{N-2}\right]
}{\sum_{\mathbb{b}}(\zb^\mathbb{b}\cdot\zb)^N}.\label{SSDCAT}
\eea
 To finish this Section, let us comment on a generalization to arbitrary $D$ of another prominent example of quantum states that are useful for quantum-enhanced 
metrology and  provide phase sensitivities beyond the standard quantum limit. We refer to Greenberger-Horne-Zeilinger (GHZ) or ``$\noon$'' (when considering bosonic particles) states. For 
$D=2$ level systems, $\noon$ states can be written in the Fock state notation \eqref{symmetricbasis} as (see e.g. \cite{RevModPhys.90.035005-Pezze})
\be
|\noon\ra=\frac{1}{\sqrt{2}}\left(|N,0\ra+e^{\ic \phi}|0,N\ra\right)=\frac{1}{\sqrt{2}}\left(\frac{(a_1^\dag)^N}{\sqrt{N!}}|\vec{0}\ra+e^{\ic \phi}\frac{(a_2^\dag)^N}{\sqrt{N!}}|\vec{0}\ra\right).\label{noonstate}
\ee
Using the canonical basis vectors $\{\eb_1,\eb_2\}$ of $\mathbb{C}^2$, we can write the $\noon$ state as a linear superposition of $\rmu(2)$ SCSs
\be
|\noon\ra=\frac{1}{\sqrt{2}}\left(e^{\ic\phi_1}|\eb_1\ra+e^{\ic\phi_2}|\eb_2\ra\right),\label{noon}
\ee
with phases $e^{\ic\phi_{1,2}}$, which coincides with \eqref{noonstate} (up to an irrelevant global phase $e^{\ic\phi_1}$) for the relative phase $\phi=\phi_2-\phi_1$. Multi-mode (or multi-level, in our context) 
generalizations of $\noon$ states have been proposed in the 
literature (see e.g. \cite{PhysRevLett.111.070403,Zhang2018}). In our scheme, this generalization of $\noon$ states 
\eqref{noon} to $D$ level systems adopts the following form
\be
|\nodon\ra=\frac{1}{\sqrt{D}}\left(\sum_{j=1}^De^{\ic\phi_j}|\eb_j\ra\right)=\frac{1}{\sqrt{D}}\left(e^{\ic\phi_1}|N,0,\dots\ra+e^{\ic\phi_2}|0,N,0,\dots\ra+\dots+e^{\ic\phi_D}|0,0,\dots,N\ra\right).\label{nodon}
\ee
EVs of linear and quadratic powers of $D$-spin operators in $\nodon$ states can be easily calculated as
\be
\la\nodon |S_{ij}|\nodon\ra=\frac{N}{D}\delta_{ij}, \quad \langle \nodon|S_{ij}S_{kl}|\nodon\rangle=\frac{N}{D}\delta_{il}\left(\delta_{jk}+(N-1)\delta_{ik}\delta_{ij}\right).\label{SNODON}
\ee

The computations in this section will be necessary to discuss entanglement and squeezing properties of all these states in the Sections \ref{entangsec} and \ref{squeezingsec}.

\section{Entanglement measures in multi-quDit systems}\label{entangsec}

In this Section we define several types of bipartition of the whole system, computing the corresponding RDMs and entanglement measures for different kinds of symmetric multi-quDit 
states $\psi$ in terms of linear $\mathcal{L}$ and von Neumann $\mathcal{S}$ entropies. We start computing interlevel entanglement in Section \ref{levelentangsec} and then (one- and two-quDit) 
interparticle entanglement  in Section \ref{partentangsec}.

\subsection{Entanglement among levels}\label{levelentangsec}

\begin{figure}[h]
\begin{center}
\includegraphics[width=7cm]{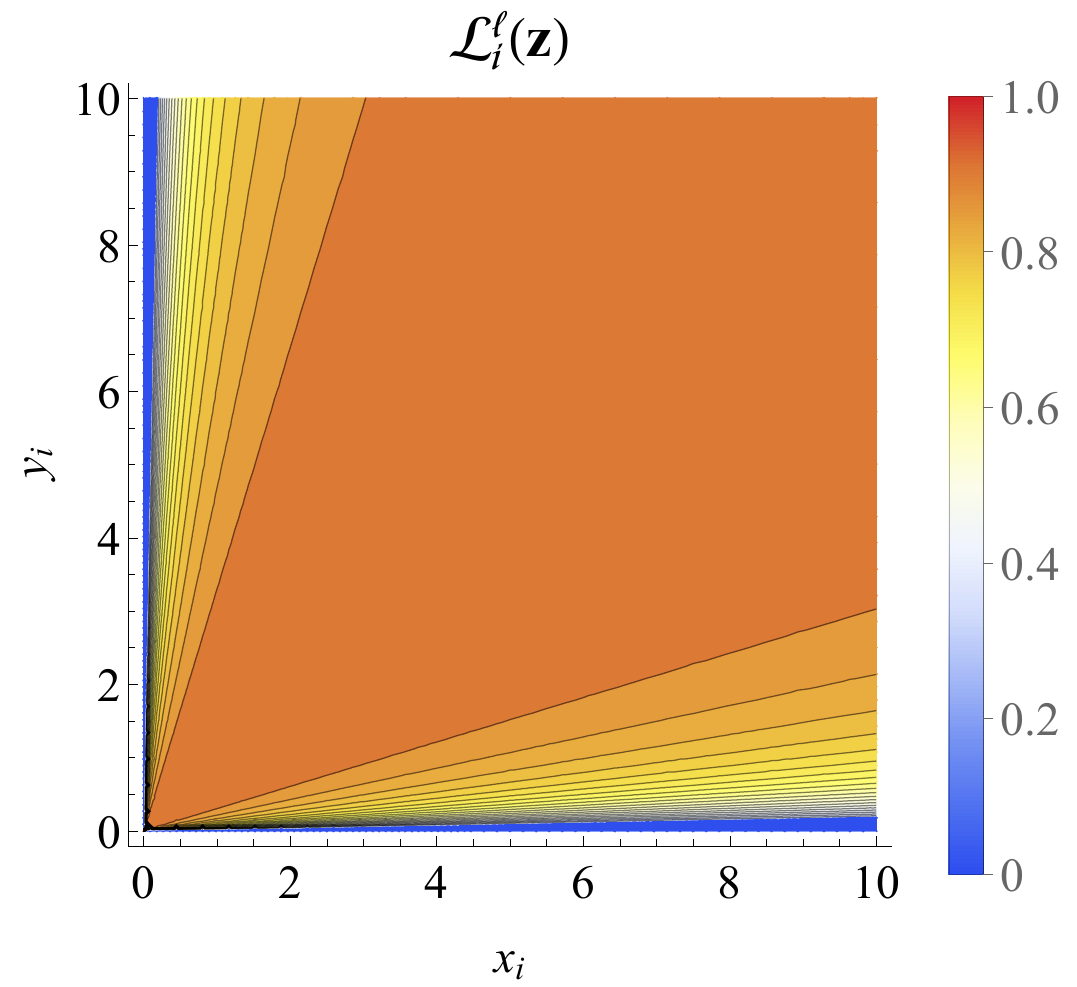}\qquad\includegraphics[width=7cm]{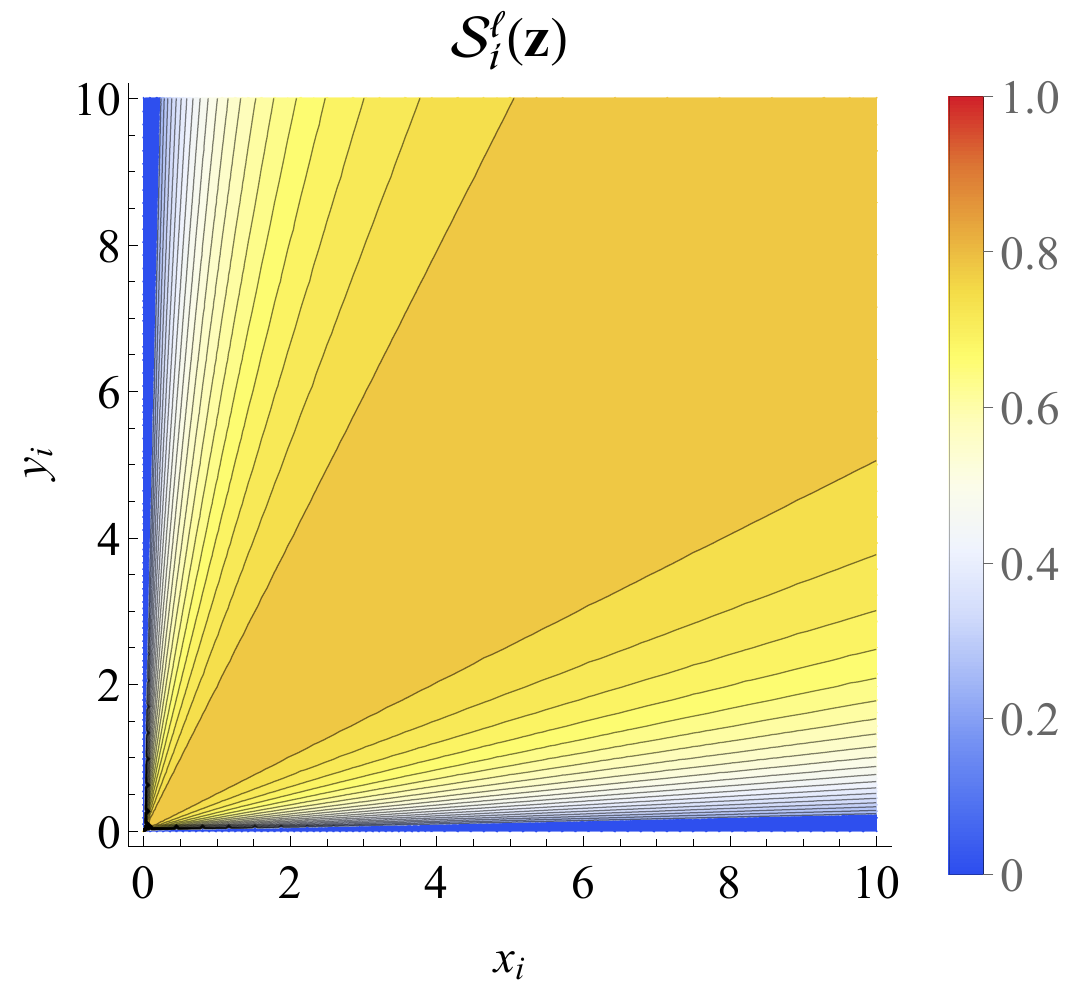}
\end{center}
\caption{Contour plots of the linear $\mathcal{L}_i^\ell$  and von Neumann $\mathcal{S}_i^\ell$ entanglement entropies, associated to the RDM of  a $\rmu(D)$-spin coherent state  $|\zb\ra$ 
of $N=10$ quDits on level $i$ \eqref{RDMleveli}, as a function of the phase-space coordinates $(x,y)$.}
\label{puritylevelCS}
\end{figure}

\begin{figure}[h]
\begin{center}
\includegraphics[width=6.5cm]{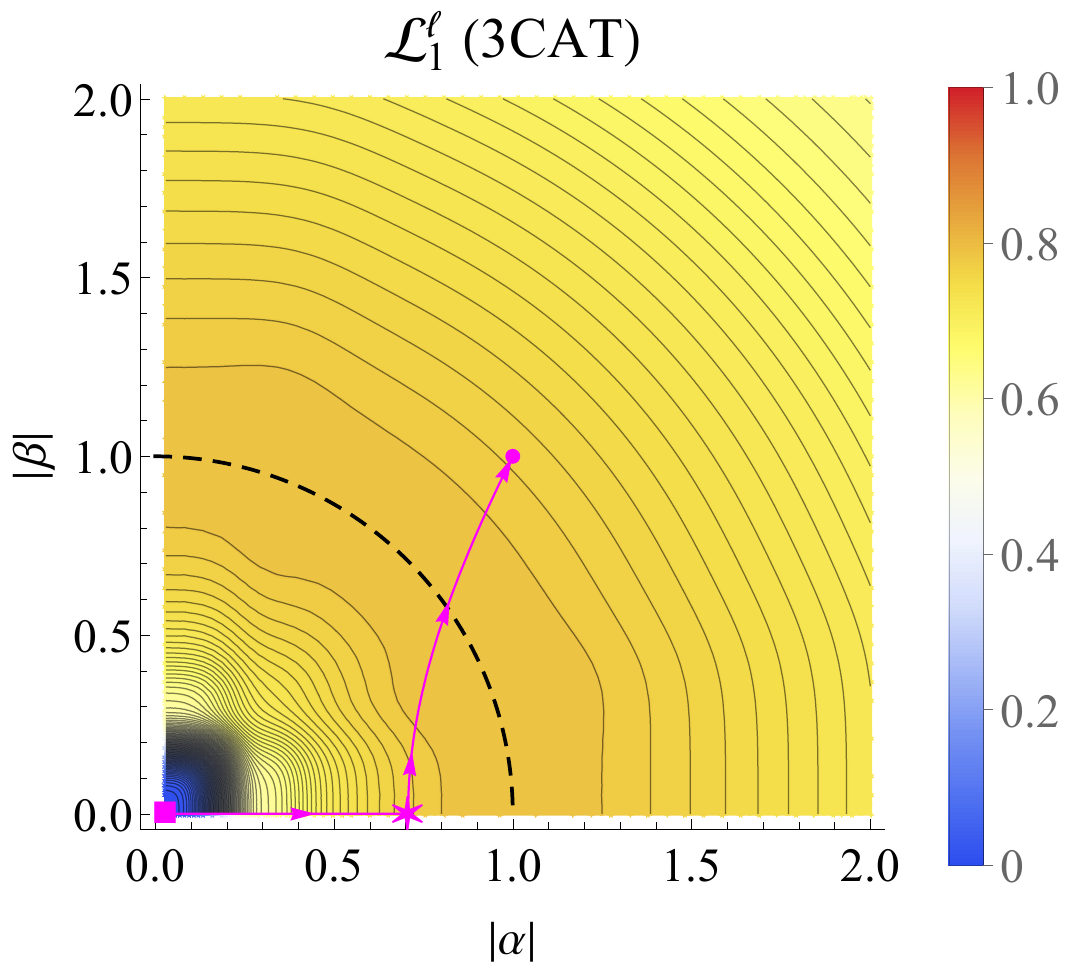}\qquad\includegraphics[width=6.5cm]{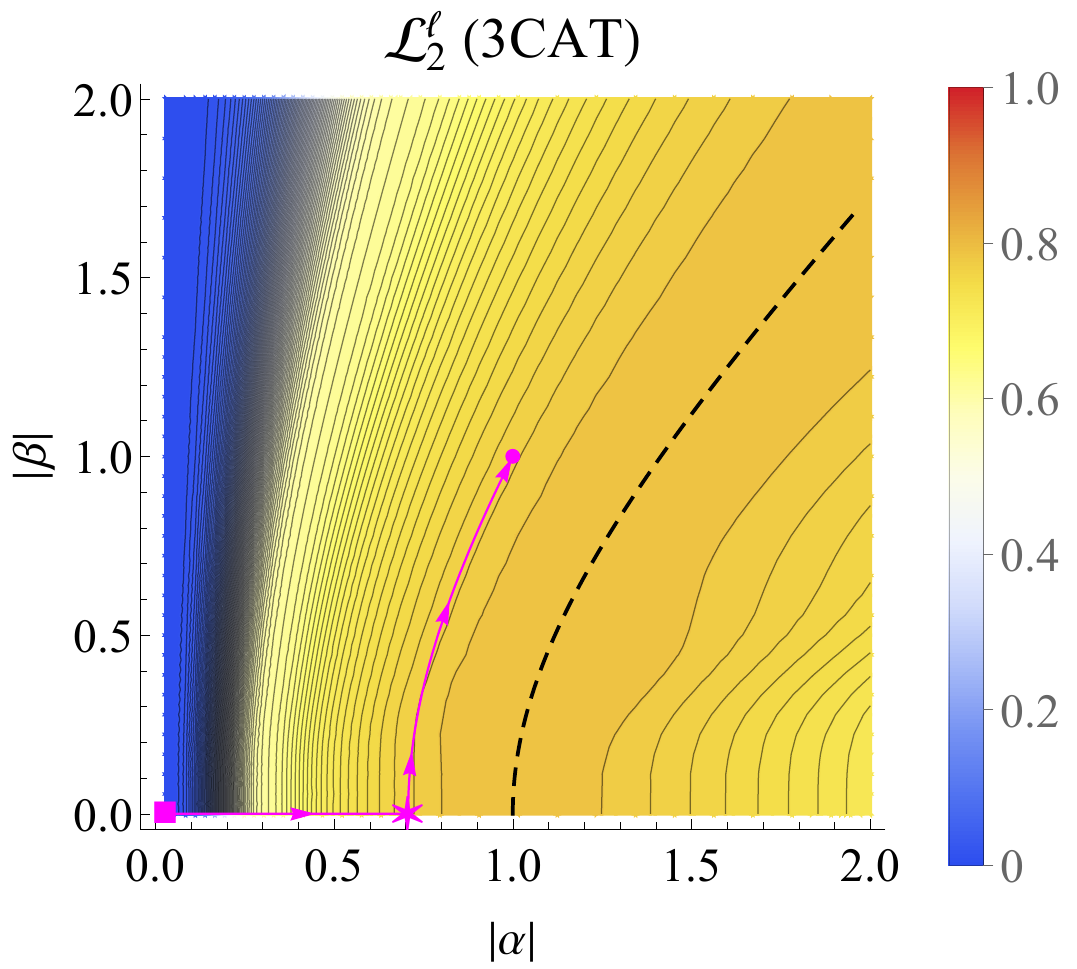}\\
\includegraphics[width=6.5cm]{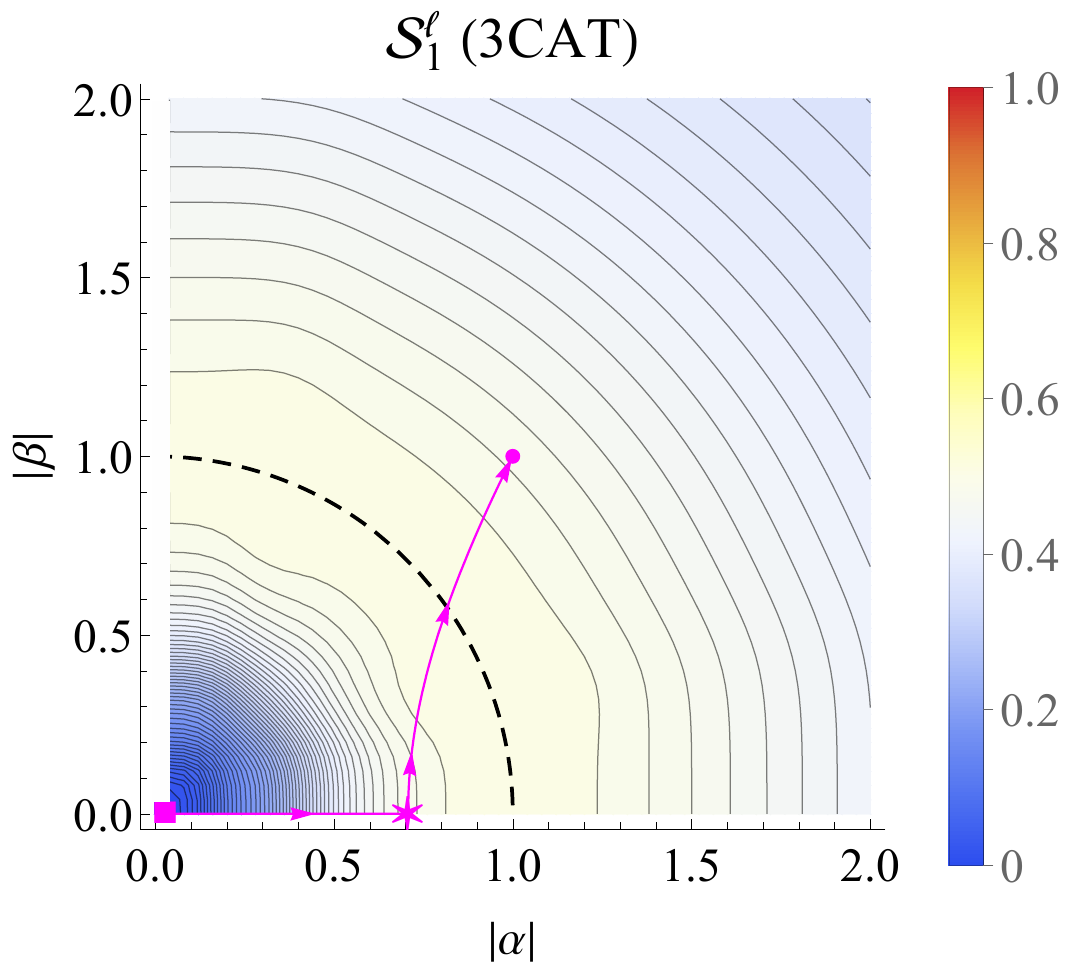}\qquad\includegraphics[width=6.5cm]{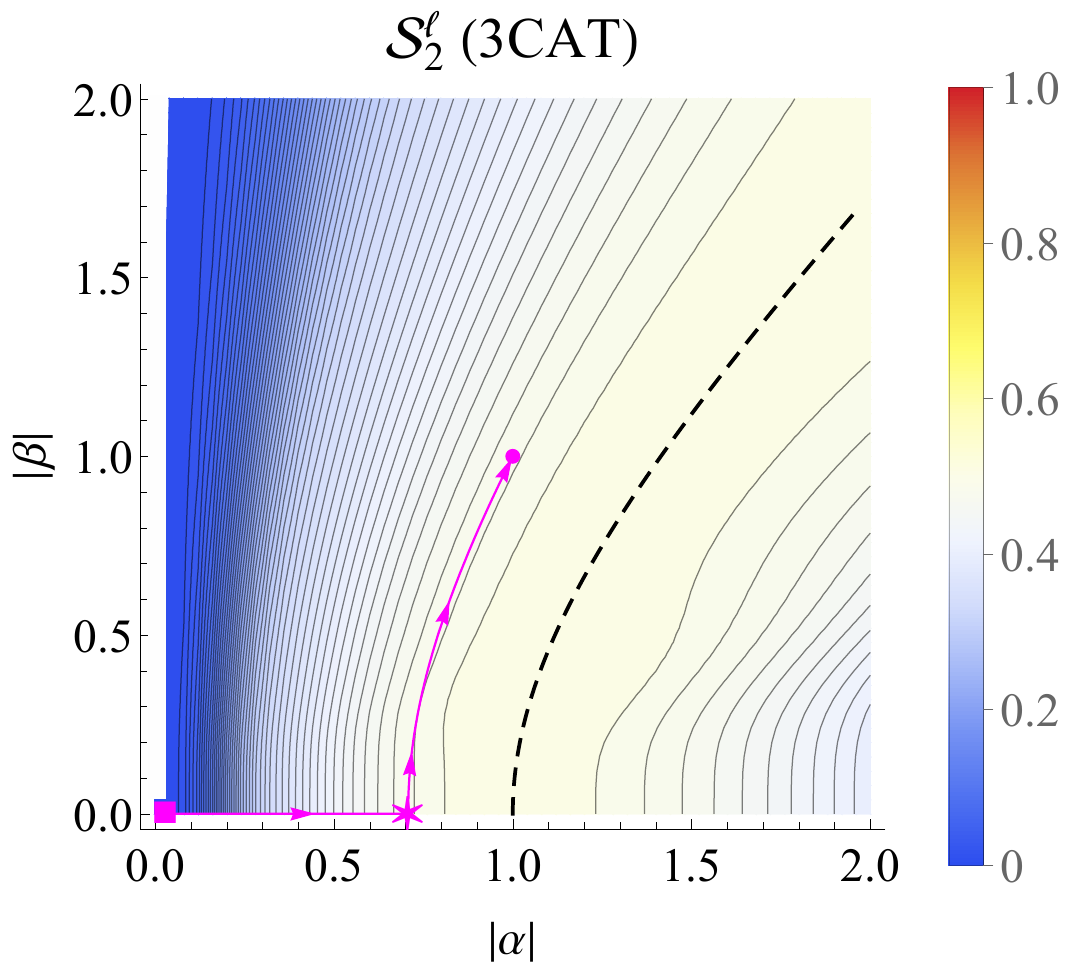}
\end{center}
\caption{Contour plots of the linear  $\mathcal{L}_i^\ell$ and von Neumann  $\mathcal{S}_i^\ell$ entanglement entropies, associated to the RDM of  a $\3cat$
of $N=20$ qutrits on levels $i=1$ and $i=2$ (the case $i=3$ is just the reflection in a diagonal mirror line of $i=2$),  
as a function of the phase-space coordinates $(\alpha,\beta)$ (they just depend on moduli). Dashed contours represent maximum entanglement entropy. The meaning of the magenta curve is explained in the main text. }
\label{puritylevel3CAT}
\end{figure}

For a general symmetric $N$-particle state $\psi$ like \eqref{psisym},  the RDM on the level $i$  is
\be
\varrho_i(\psi)=\tr_{j\neq i}\Bigg(\sum_{\vec{n},\vec{n}'}{}'\,c_{\vec{n}'}\bar{c}_{\vec{n}}|n'_1,\dots, n'_D\ra\la n_1,\dots, n_D|\Bigg)=\sum_{\vec{n}}{}'\,|c_{\vec{n}}|^2|n_i\ra\la n_i|.
\ee
Thus $\varrho_i(\psi)$ lies in a single boson Hilbert space of dimension $N+1$. Its purity  is then 
\be 
\mathcal{P}_i^\ell(\psi)=\tr(\varrho_i^2(\psi))=\sum_{\vec{n},\vec{m}}{}'\,|c_{\vec{n}}|^2|c_{\vec{m}}|^2\delta_{n_i,m_i}.\label{purylevel}
\ee
Here the superscript $\ell$ makes reference to ``level'', to distinguish it from ``atom'' purity $\mathcal{P}^\mathrm{a}$ in the next section. It can also make reference to entanglement 
between different boson species $\ell$, like rotational-vibrational entanglement \cite{Calixto_2012,PhysRevA.89.032126} in algebraic molecular models \cite{Iachellolevine,Frankvanisacker} 
such as the vibron model based on a bosonic $\rmu(3)$ spectrum-generating algebra \cite{PhysRevA.77.032115,PhysRevA.83.062125}. 
For the case of the DSCS  $|\zb\ra$ in \eqref{cohND}, taking the coefficients  $c_\nb$ in \eqref{coefCS}, and 
after a lengthy calculation, the  RDM on level $i$ turns out to be diagonal
\be
\varrho_i(\zb)=\sum_{n=0}^N \lambda_n(x_i,y_i) |N-n\ra\la N-n|,\quad  \lambda_n(x_i,y_i)=\binom{N}{n}\frac{x_i^{N-n}y_i^{n}}{(x_i+y_i)^{N}},\quad  x_i=|z_i|^2, y_i=|\zb|^2-|z_i|^2.\label{RDMleveli}
\ee
Note that the eigenvalues $\lambda_n$ can be expressed in terms of only two positive real coordinates $(x_i,y_i)$, except for the reference level $z_i=1$, for which $x_i=1$ 
and therefore there is only one independent variable $y_i=|\zb|^2-1$. For example, for $\rmu(3)$ SCSs, choosing $i=1$ as the reference level and  using 
the parametrization $\zb=(1,\alpha,\beta)$ for the phase space $\mathbb{C}P^2$ in \eqref{S3C}, we have $x_1=1, x_2=|\alpha|^2, x_3=|\beta|^2$ and $y_1=|\alpha|^2+|\beta|^2, y_2=1+|\beta|^2, y_3=1+|\alpha|^2$. 
The purity of $\varrho_i(\zb)$ is simply $\mathcal{P}_i^\ell(x_i,y_i)=\sum_{n=0}^N \lambda_n^2(x_i,y_i)$.  In Figure \ref{puritylevelCS} we represent the Linear and von Neumann 
\be \mathcal{L}_i^\ell=\frac{N+1}{N}(1-\mathcal{P}_i^\ell), \quad   \mathcal{S}_i^\ell=-\sum_{n=0}^N \lambda_n\log_{N+1}\lambda_n
\ee
entanglement entropies for a general level $i$ as a function of $(x,y)$ [for the reference level $i=1$, we have to restrict ourselves to the cross section $x=1$]. We normalize linear and von Neumann entropies 
so that their interval range is $[0,1]$, the extremal values corresponding to pure and completely mixed RDMs, respectively. We shall see that both entropies, $\mathcal{L}$ and $\mathcal{S}$, 
provide similar qualitative behavior for the bipartitions studied in this paper. 
Interlevel isentropic contours correspond to the straight lines $y=mx$ (see Figure \ref{puritylevelCS}), and the maximum is attained for $y=x$. The large 
$N$ behavior of the interlevel linear entanglement entropy $\mathcal{L}_i^\ell(\zb)$ around the maximum $y=x$ is  $\mathcal{L}_i^\ell=1-1/\sqrt{\pi N}+O(N^{-3/2})$. 
In Figure \ref{puritylevel3CAT} we represent contour plots of $\mathcal{L}_{1,2}^\ell$ and $\mathcal{S}_{1,2}^\ell$ for the RDM  of a  
$\3cat$ of $N=20$ qutrits on levels $i=1$ and $i=2$. Note that linear and von Neumann entropies display a similar structure. 
We omit $\mathcal{L}_3^\ell$ and $\mathcal{S}_3^\ell$ since they are just the reflection in a diagonal mirror line of $\mathcal{L}_2^\ell$ and $\mathcal{S}_2^\ell$, respectively.  
$\mathcal{L}_1^\ell$ attains its maximum at the isentropic circle $|\alpha|^2+|\beta|^2=1$, whereas  $\mathcal{L}_2^\ell$ attains its maximum at the isentropic hyperbola $|\alpha|^2-|\beta|^2=1$. 
The large  $N$ behavior of the interlevel linear entanglement entropy for a $\dcat$  around the maximum  is  $\mathcal{L}_i^\ell=1-2/\sqrt{\pi N}+O(N^{-3/2})$. Figure \ref{puritylevel3CAT} also shows (in magenta color) the parametric curve 
$(\alpha(\lambda),\beta(\lambda))$ obtained later in Section \ref{LMGsec} and related to the stationary points \eqref{critalphabeta} of the energy surface \eqref{enersym} in the quantum phase diagram of a LMG 3-level atom model, where $\lambda$ is the atom-atom interaction coupling constant. 
For high interactions we have $(\alpha(\lambda),\beta(\lambda))\stackrel{\lambda\to\infty}{\longrightarrow} (1,1)$, 
which does not lie inside the maximum isentropic curve, although the difference between both entropies tends to zero in the limit $N\to\infty$ (see later in Sections \ref{LMGsec} and \ref{signaturesec} for more information).

For $\nodon$ states \eqref{nodon}, the RDM on level $i$ and its purity are 
\be
\varrho_i(\nodon)=\frac{D-1}{D}|0\ra_i\la 0|+\frac{1}{D}|N\ra_i\la N|, \quad \tr(\varrho_i^2(\nodon))=1-2\left(\frac{D-1}{D^2}\right),
\ee
which is independent of the level $i$. Therefore, the linear entropy is given by $\mathcal{L}^\ell_i(\nodon)= 2\frac{N+1}{N}\frac{D-1}{D^2}$, which reduces to $\mathcal{L}^\ell_i(\noon)=\frac{N+1}{2N}$ for $D=2$.

\subsection{Entanglement among atoms}\label{partentangsec}

We compute the one- and  two-particle RDMs for a single and a pair of particles extracted at random from a symmetric $N$-quDit state. The corresponding entanglement entropies 
are expressed in terms of EVs of collective $D$-spin operators $S_{ij}$. 

\subsubsection{One-quDit reduced density matrices}

\begin{figure}[h]
\begin{center}
\includegraphics[width=7cm]{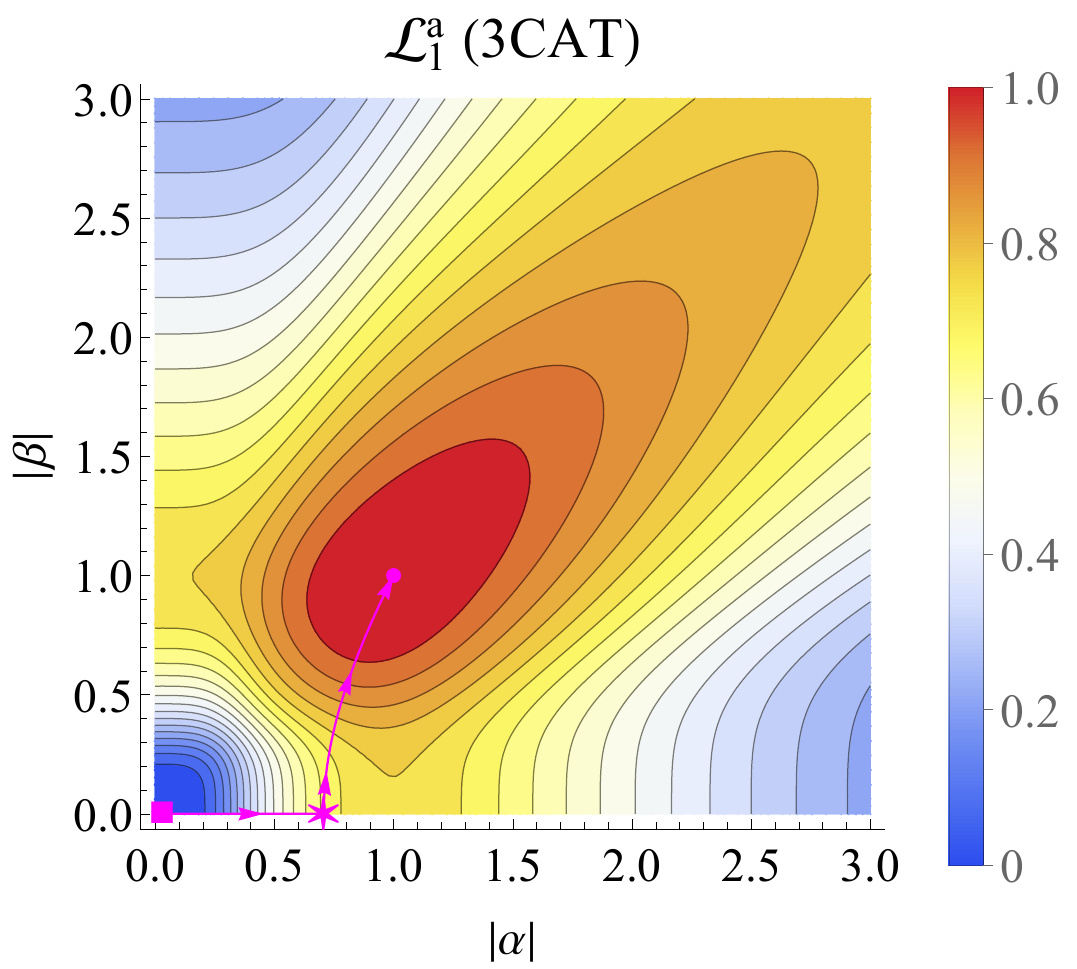}\qquad\includegraphics[width=7cm]{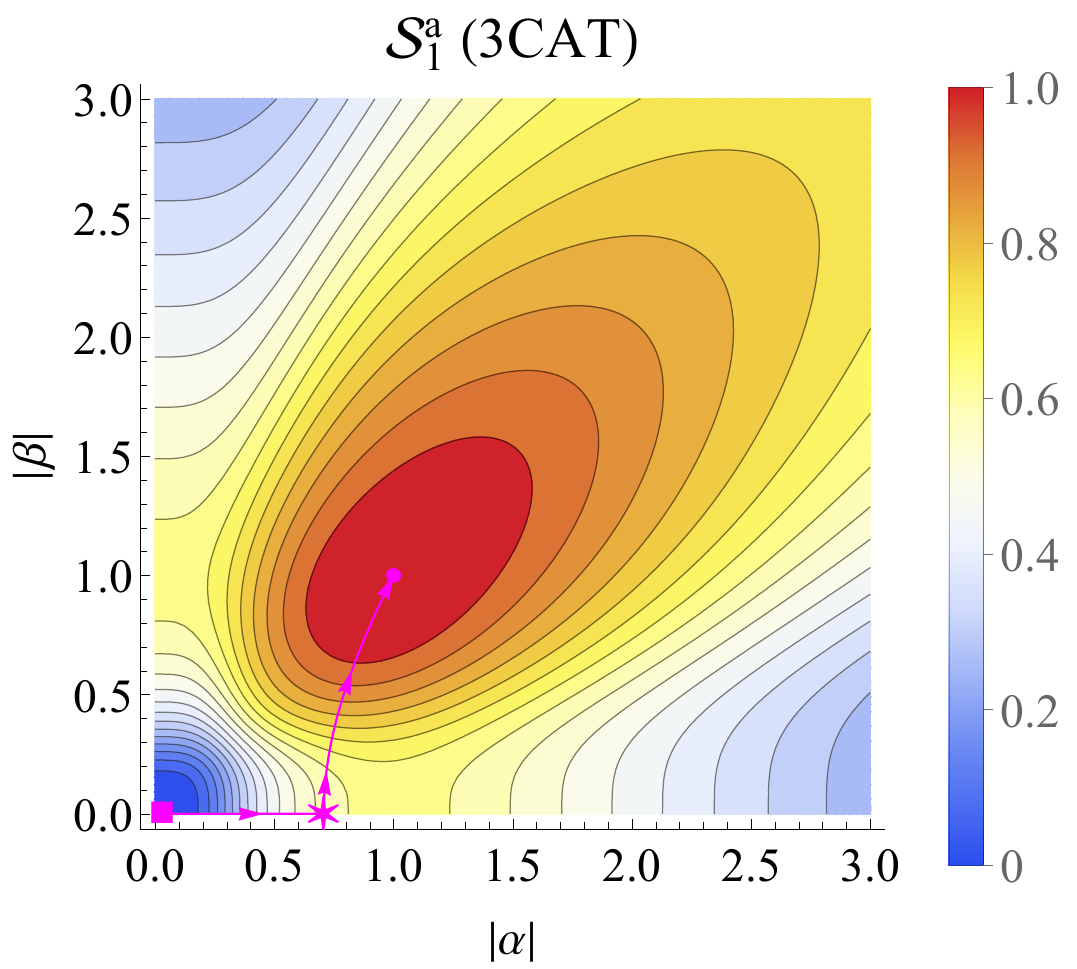}
\end{center}
\caption{3D plots of linear $\mathcal{L}_1^\mathrm{a}$ and von Neumann $\mathcal{S}_1^\mathrm{a}$ entanglement entropies of the one-qutrit RDM $\rho_1(\3cat)$ of a $\rmu(3)$ Schr\"odinger cat \eqref{S3C} for $N=10$ atoms, as a function 
of the phase-space coordinates $\alpha, \beta$ (they just depend on moduli). The meaning of the magenta curve is the same as in the Figure \ref{puritylevel3CAT}.}
\label{purityOne3CAT}
\end{figure}

Any density matrix of a single quDit  can be written as a combination of Hubbard matrices $\E_{ij}$ with commutation relations \eqref{commurel}   as
\be
\rho_1=\sum_{i,j=1}^D r_{ij}\E_{ij}, \quad r_{ij}=\tr(\rho_1\E_{ji})=\la \E_{ji}\ra
\ee
with $r_{ij}$ complex numbers (the generalized Bloch vector) fulfilling $\bar{r}_{ij}=r_{ji}$ and (the generalized Bloch sphere) 
\be \tr[\rho_1]=\sum_{i=1}^D r_{ii}=1, \quad  0<\tr[(\rho_1)^2]=\sum_{i,j=1}^D |r_{ij}|^2\leq 1.
\ee
We want to construct the one-quDit RDM for one quDit extracted at random from a symmetric $N$-quDit state $\psi$. 
The procedure consists of writing the one-quDit RDM entries in terms of expectation values (EVs) of collective  $D$-spin operators \eqref{collectiveS}. Remember the definition of $E_{ij}^\mu$, $\mu=1,\dots,N$ after \eqref{commurel} as 
the embedding of the single $\mu$-th atom  $E_{ij}$ operator into the $N$-atom Hilbert space. Atom indistinguishableness implies that $\la {\E}_{ij}^\mu\ra=\frac{1}{N}\la S_{ij}\ra$, for any $\mu=1,\dots, N$, and therefore the 
one-quDit RDM of any normalized symmetric $N$-quDit state $\psi$ like  \eqref{psisym} can be  expressed as 
\be \rho_{1}(\psi)=\frac{1}{N}\sum_{i,j=1}^D \la S_{ji}\ra \E_{ij}.\ee
with  $D$-spin EVs \eqref{isoEV}. Note that $\tr[\rho_{1}(\psi)]=1$ since $\sum_{i=1}^D\la S_{ii}\ra=N$ (total population of the $D$ levels), which is related to the  linear Casimir operator 
$C_1=\sum_{i=1}^DS_{ii}$ of $\rmu(D)$. From the condition $\tr(\rho_{1}(\psi)^2)\leq 1$ we obtain the general relation
\be
\sum_{i,j=1}^D|\la S_{ij}\ra|^2\leq N^2,\label{S2}
\ee
which could be seen as a measure of the fluctuations or departure from the linear $C_1$ and quadratic $C_2$  Casimir operators given by $C_1=N\mathbb{1}$ and 
\be
C_2=\sum_{i,j=1}^D S_{ij}S_{ji}=N(N+D-1)\mathbb{1}.\label{Casimires}
\ee
The quantum limit $N^2$ in  \eqref{S2} is attained  for DSCSs. Indeed, for $|\psi\ra=|\zb\ra$ in \eqref{cohND}, the DSCS operator EVs were calculated in \eqref{CSEV}. Therefore, the purity of 
the corresponding one-quDit/atom RDM is simply [we denote interatom purity by $\mathcal{P}^\mathrm{a}$ to distinguish from the interlevel 
purity $\mathcal{P}^\ell$ discussed in the previous section]
\be
\mathcal{P}_1^\mathrm{a}(\zb)=\tr(\rho_{1}(\zb)^2)=\frac{1}{N^2}\sum_{i,j=1}^D|\la\zb|S_{ij}|\zb\ra|^2=\sum_{i,j=1}^D \frac{|z_i|^2|z_j|^2}{|\zb|^4}=1,\ee
which means that there is not entanglement between atoms in a DSCS . This is because a DSCS is eventually obtained by rotating each atom individually. The situation 
changes when we deal with parity adapted DSCSs or ``Schr\"odinger cat states'' \eqref{SC}. Indeed, the one-quDit RDM $\rho_1(\dcat)$ does not correspond 
now to a pure state since, using the $D$-spin EVs on a $\dcat$ state \eqref{SDCAT}, the 
purity gives
\be
\mathcal{P}_1^\mathrm{a}(\dcat)=\tr(\rho_{1}(\dcat)^2)=\frac{\left(\sum_{\mathbb{b}}(\zb^\mathbb{b}\cdot\zb)^{N-1}\right)^2+
\sum_{i=2}^D|z_i|^4\left(\sum_{\mathbb{b}}(-1)^{b_i}(\zb^\mathbb{b}\cdot\zb)^{N-1}\right)^2}{\left(\sum_{\mathbb{b}}(\zb^\mathbb{b}\cdot\zb)^N\right)^2} \leq 1.\label{purityDCAT}
\ee
That is, unlike $|\zb\ra$, the Schr\"odinger cat $|\dcat\ra$ is not separable in the tensor product Hilbert space  $[\mathbb{C}^D]^{\otimes N}$. In Figure \ref{purityOne3CAT}, we 
represent contour  plots of linear  and von Neumann 
\be
\mathcal{L}_1^\mathrm{a}=\frac{D}{D-1}(1-\mathcal{P}_1^\mathrm{a}),\quad \mathcal{S}_1^\mathrm{a}=-\tr(\rho_{1}\log_D \rho_{1})
\ee
entanglement entropies of the one-qutrit RDM $\rho_1(\3cat)$ of a $\rmu(3)$ Schr\"odinger cat \eqref{S3C} as a function 
of the phase-space $\mathbb CP^{2}$ coordinates $\alpha, \beta$ [actually, they just depend on the moduli]. Both entropies are again normalized to 1.  They attain their maximum 
value of 1 at the phase-space point $(\alpha,\beta)=(1,1)$ corresponding to a maximally mixed RDM. Figure \ref{purityOne3CAT} also shows (in magenta color) the stationary curve 
$(\alpha(\lambda),\beta(\lambda))$  previously mentioned at the end of Sec. \ref{levelentangsec} in relation to the Figure \ref{puritylevel3CAT}. 
For high interactions we have $(\alpha(\lambda),\beta(\lambda))\stackrel{\lambda\to\infty}{\longrightarrow} (1,1)$, which means that highly coupled atoms are maximally entangled in a cat-like ground state 
(see later in Sections \ref{LMGsec} and \ref{signaturesec} for more information).

To finish this Section, let us comment on one-quDit entanglement for $\nodon$ states \eqref{nodon}. Taking into account the $D$-spin EV \eqref{SNODON}, the one-quDit RDM of a $\nodon$ 
is simply $\rho_{1}(\nodon)=\frac{1}{D}\mathbb{1}_D$ 
and its linear entropy   $\mathcal{L}_1^\mathrm{a}=1$, implying maximally mixed RDM.

\subsubsection{Two-quDit reduced density matrices}

Likewise, any density matrix of  two quDits can be written as
\be
\rho_2=\sum_{i,j,k,l=1}^D r_{ijkl}\E_{ij}\otimes\E_{kl},\quad r_{ijkl}=\tr(\rho_2\E_{ji}\otimes\E_{lk})=\la \E_{ji}\otimes\E_{lk}\ra.
\ee
with $\bar{r}_{ijkl}={r}_{jilk}$ complex  parameters subject to   $\tr[\rho_2]=1$ and $0<\tr[(\rho_2)^2]\leq 1$. 
Now we need to express the RDM on  a pair of particles, extracted at random from a symmetric state of $N$ 
$D$-level atoms, in terms of EVs of bilinear products of collective $D$-spin operators $S$. In particular, we have
\bea
\la S_{ij} S_{kl}\ra&=&\sum_{\mu,\nu=1}^N \la {\E}_{ij}^\mu{\E}_{kl}^\nu\ra =\sum_{\mu=1}^N \delta_{jk}\la{\E}_{il}^\mu\ra+\sum_{\mu\not=\nu=1}^N \la {\E}_{ij}^\mu{\E}_{kl}^\nu\ra\nn\\ 
&=&  \delta_{jk}\la S_{il}\ra+N(N-1) \la {\E}_{ij}^1{\E}_{kl}^2\ra,
\eea
due to indistinguishableness. Therefore, the two-particle RDM of a symmetric state $\psi$  of $N>2$ quDits is written as
\be
\rho_2(\psi)=\frac{1}{N(N-1)}\sum_{i,j,k,l=1}^D (\la S_{ji} S_{lk}\ra-\delta_{il}\la S_{jk}\ra)  \E_{ij}\otimes\E_{kl}.\label{rho2}
\ee
Using the Casimir values \eqref{Casimires}, one can directly prove  that  $\tr[\rho_2(\psi)]=1$ for any normalized symmetric state $\psi$. 
The case $D=2$ was considered by Wang and M\o{}lmer in \cite{Molmer}. The procedure is straightforwardly extended to $\rho_M$ for an arbitrary number $M\leq N/2$ of quDits. 
The purity of $\rho_2(\psi)$ can be compactly written as
\begin{align}
\mathcal{P}_2^\mathrm{a}(\psi)=\tr{\left(\rho_2(\psi)^2\right)}=\frac{1}{N^2(N-1)^2}\left[\sum_{i,j,k,l=1}^D \braket{S_{ji}S_{lk}}\braket{S_{ij}S_{kl}}-2\sum_{i,j,k=1}^D \braket{S_{ji}S_{kj}}\braket{S_{ik}}+\sum_{i,j=1}^D 
\braket{S_{ii}}\braket{S_{jj}}\right].\label{pur2qudit}
\end{align}

In order to construct the two-particle RDM of a DSCS \eqref{cohND}, we need the EVs of quadratic powers \eqref{qEV}. 
With these ingredients, we can easily compute the two-particle RDM of a DSCS \eqref{cohND} which, for large $N$ has the following asymptotic expression
\be
\rho_{2}(\zb)=\sum_{i,j,k,l=1}^D \left(\bar{z}_jz_i\bar{z}_lz_k+O(1/N)\right) \E_{ij}\otimes\E_{kl}.
\ee
The purity of $\rho_{2}(\zb)$ is 1 since $\zb$ is separable in the tensor product Hilbert space $[\mathbb{C}^D]^{\otimes N}$, as we have already commented. Moreover, one can see 
that $\rho_{2}(\zb)=\rho_{1}(\zb)\otimes \rho_{1}(\zb)$. 
However, the Schr\"odinger cat \eqref{SC} is non-separable and has an intrinsic pairwise entanglement. Taking into account the particular estructure of linear \eqref{SDCAT} and quadratic \eqref{SSDCAT} $D$-spin operator EVs, 
The general formula \eqref{pur2qudit} becomes 

\begin{align}
 \mathcal{P}_2^\mathrm{a}(\dcat)=\frac{1}{N^2(N-1)^2}\left[\sum_{\substack{i,j,k,l=1\\ j\neq k}}^D 
    \braket{S_{ji}S_{lk}}\braket{S_{ij}S_{kl}}+\sum_{i,j=1}^D \braket{S_{ji}S_{ij}}\big(\braket{S_{ij}S_{ji}}-2\braket{S_{ii}}\big)+\sum_{i,j=1}^D \braket{S_{ii}}\braket{S_{jj}}\right].
\end{align}

\begin{figure}[h]
\begin{center}
\includegraphics[width=7cm]{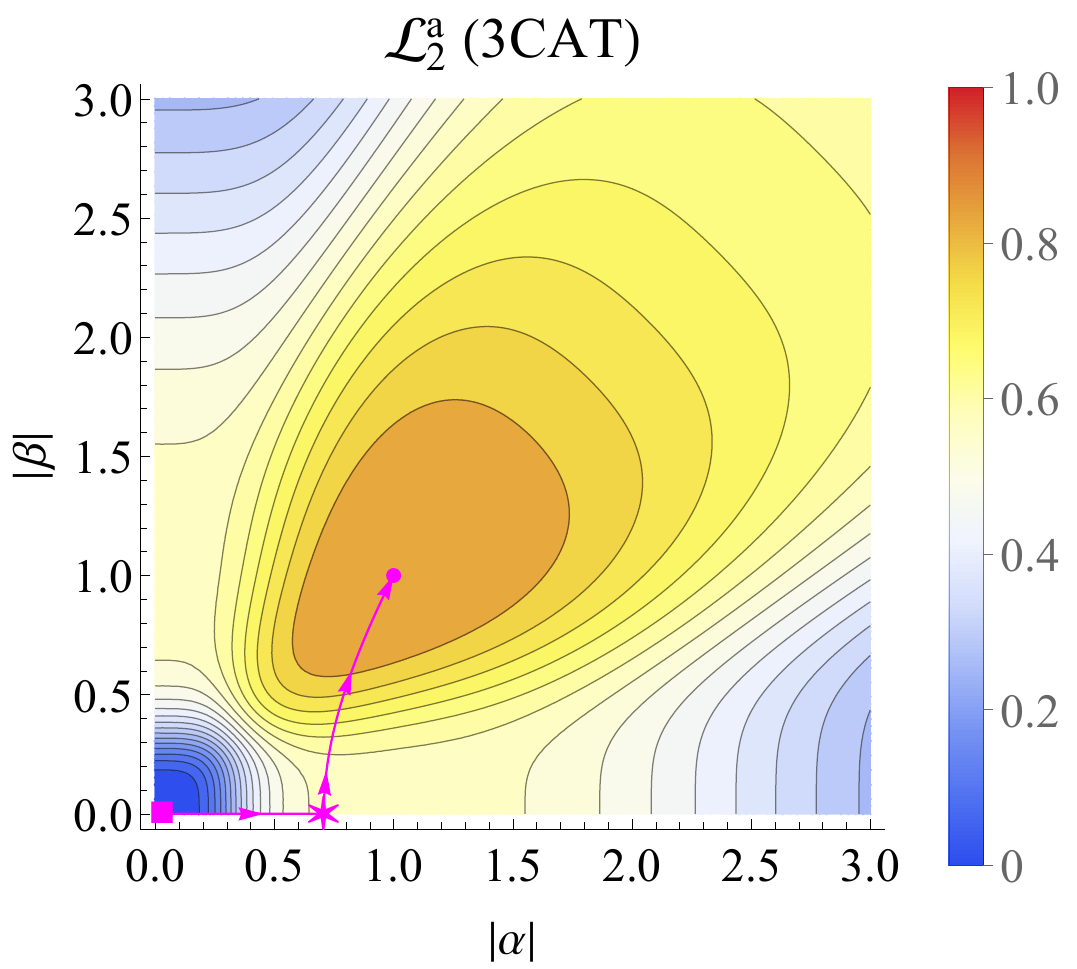}\qquad\includegraphics[width=7cm]{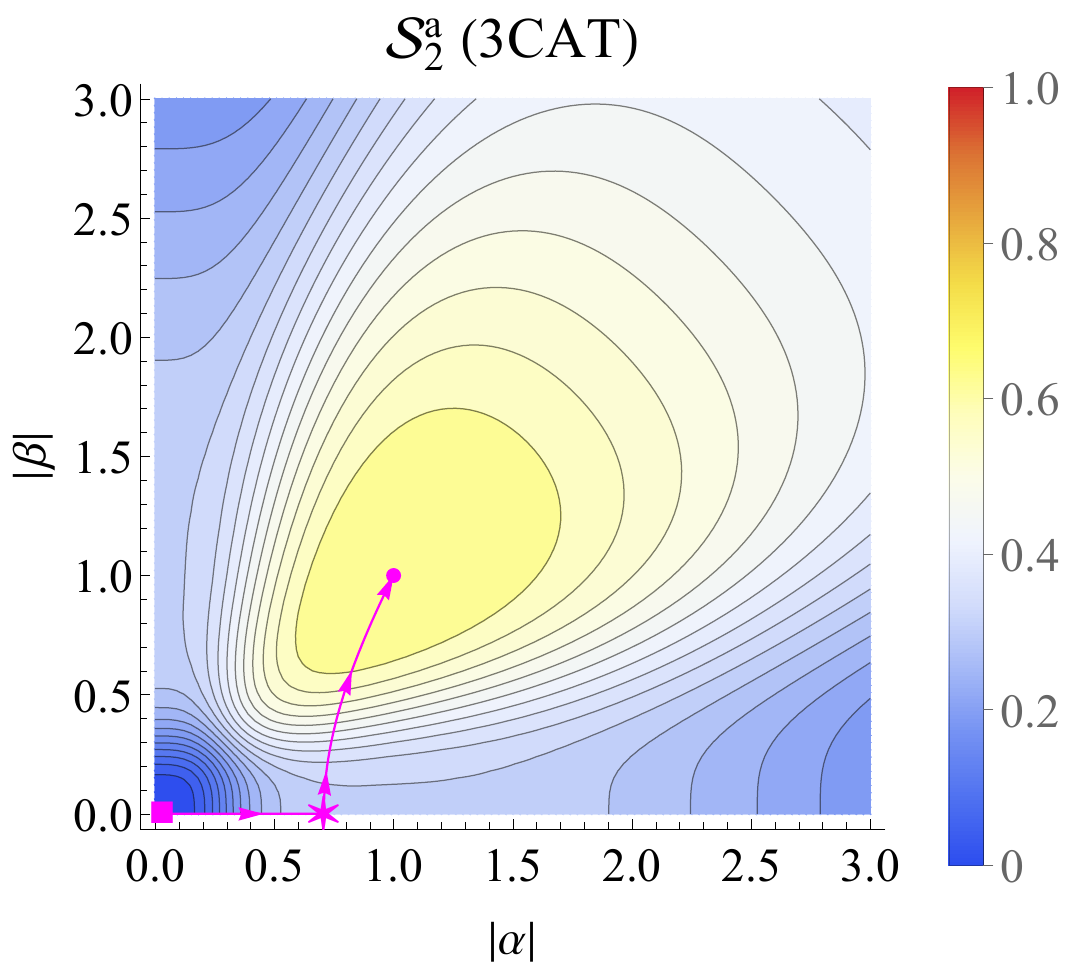}
\end{center}
\caption{Contour plots of linear $\mathcal{L}_2^\mathrm{a}$ and von Neumann $\mathcal{S}_2^\mathrm{a}$ entanglement entropies of the two-qutrit RDM $\rho_2(\3cat)$ of a $\rmu(3)$ Schr\"odinger cat \eqref{S3C} for $N=10$ atoms, as a function 
of the phase-space coordinates $\alpha, \beta$ (they just depend on moduli). The meaning of the magenta curve is the same as in the Figure \ref{puritylevel3CAT}.}
\label{purityTwo3CAT}
\end{figure}
In Figure \ref{purityTwo3CAT}, we 
represent contour plots of normalized linear  and von Neumann 
\be
\mathcal{L}_2^\mathrm{a}=\frac{D^2}{D^2-1}(1-\mathcal{P}_2^\mathrm{a}),\quad \mathcal{S}_2^\mathrm{a}=-\tr(\rho_{2}\log_{D^2} \rho_{2})
\ee
entanglement entropies for the two-qutrit RDM $\rho_2(\3cat)$ of a $\rmu(3)$ Schr\"odinger cat \eqref{S3C} as a function 
of the phase-space $\mathbb CP^{2}$ coordinates $\alpha, \beta$ [they just depend on the moduli]. As for the one-quDit case, they attain their maximum 
value at the phase-space point $(\alpha,\beta)=(1,1)$; however, unlike the one-quDit case, pairwise entanglement entropies do not attain the maximum value of 1 at this point, but 
$\mathcal{L}_2^\mathrm{a}=5/6$ and $\mathcal{S}_2^\mathrm{a}\simeq 0.623$ for large $N$. 
As already commented, variational (spin coherent) approximations to the ground state 
of the  LMG 3-level atom model [discussed later in Section  \ref{LMGsec}] recover this maximum entanglement point  $(\alpha,\beta)=(1,1)$ at high interactions $\lambda\to\infty$ (as can be seen in the magenta curve).

For $\nodon$ states \eqref{nodon}, the two-quDit RDM is
\be \rho_{2}(\nodon)=\frac{1}{D} \sum_{k=1}^D \E_{kk}\otimes\E_{kk}, 
\ee
and therefore $\rho_{2}(\nodon)^2=\frac{1}{D}\rho_{2}(\nodon)$, which means that the linear entropy is $\mathcal{L}_2^\mathrm{a}(\nodon)=D/(D+1)$, indicating a high level of pairwise entanglement in a $\nodon$ state.

\section{SU(D) spin squeezing: a proposal}\label{squeezingsec}

As we have already commented in the introduction, Wang and Sanders \cite{PhysRevA.68.012101-Wang} showed a direct relation between the concurrence $C$, extracted from the two-qubit RDM \eqref{rho2} for $D=2$, and the 
$\mathrm{SU}(2)$ spin $\vec{J}=(J_x,J_y,J_z)$ squeezing parameter 
\be
\xi^2=\frac{4}{N}\min_\theta\la(\cos(\theta)J_x+\sin(\theta)J_y)^2\ra=\frac{2}{N}\left[\la J_x^2+J_y^2\ra-\sqrt{\la J_x^2-J_y^2\ra^2+\la J_xJ_y+J_yJ_x\ra^2}\right]\label{squeezingpar}
\ee
introduced by \cite{Kitagawa}, which measures spin fluctuations in an orthogonal direction to the mean value $\la \vec{J}\ra$ with minimal variance. Actually, the definition \eqref{squeezingpar} refers to 
even and odd symmetric multi-qubit states [remember the extension of this 
concept to multi-quDits after  \eqref{parityop}] for which $\la \vec{J}\ra=(0,0,\la J_z\ra)$ and therefore the orthogonal direction lies in the plane XY. This definition can be extended to even and odd symmetric multi-quDit states 
for which $\la S_{ij}\ra\propto \delta_{ij}$ [see e.g. \eqref{SDCAT} for the case of the even $\dcat$ state]. Using the embedding \eqref{su2embedding} of $D(D-1)/2$ $\mathrm{SU}(2)$ spin subalgebras into $\rmu(D)$, and 
mimicking \eqref{squeezingpar}, we can define $D(D-1)/2$ spin squeezing parameters $\xi_{ij}, i>j$ for $D$-spin systems as:
\be
\xi_{ij}^2=\frac{1}{N(D-1)}\left[ \la S_{ij}S_{ji}+S_{ji}S_{ij}\ra-2|\la S_{ij}^2\ra|  \right], \quad  i>j=1,\dots,D-1.\label{squeezingparD}
\ee
We have chosen the normalization factor $\frac{1}{N(D-1)}$ so that \eqref{squeezingparD} reduces to \eqref{squeezingpar} for $D=2$ and so that the total $D$-spin squeezing parameter
\be
\xi^2_D=\sum_{i>j=1}^D \xi_{ij}^2\label{totalsqueezing}
\ee
is one (no squeezing) for the DSCSs $|\zb\ra$ in \eqref{cohND}. Actually, for 
DSCSs we have that $\xi_{ij}^2=(|z_i|^2+|z_j|^2)/(|\zb|^2(D-1))$ is written in terms of average level populations $\la\zb|S_{ii}|\zb\ra=N|z_i|^2/|\zb|^2$ of levels $i$ and $j$,  
acording to \eqref{CSEV}. Therefore, the presence of $D$-spin 
squeezing means in general that  $\xi^2_D<1$. Using the EVs \eqref{SNODON} for $\nodon$ states \eqref{nodon}, the 
corresponding spin squeezing parameters are $\xi_{ij}=2/[D(D-1)]$, which gives $\xi^2_D=1$, thus implying that $\nodon$ states do not exhibit spin squeezing.

\begin{figure}[h]
\begin{center}
\includegraphics[width=7cm]{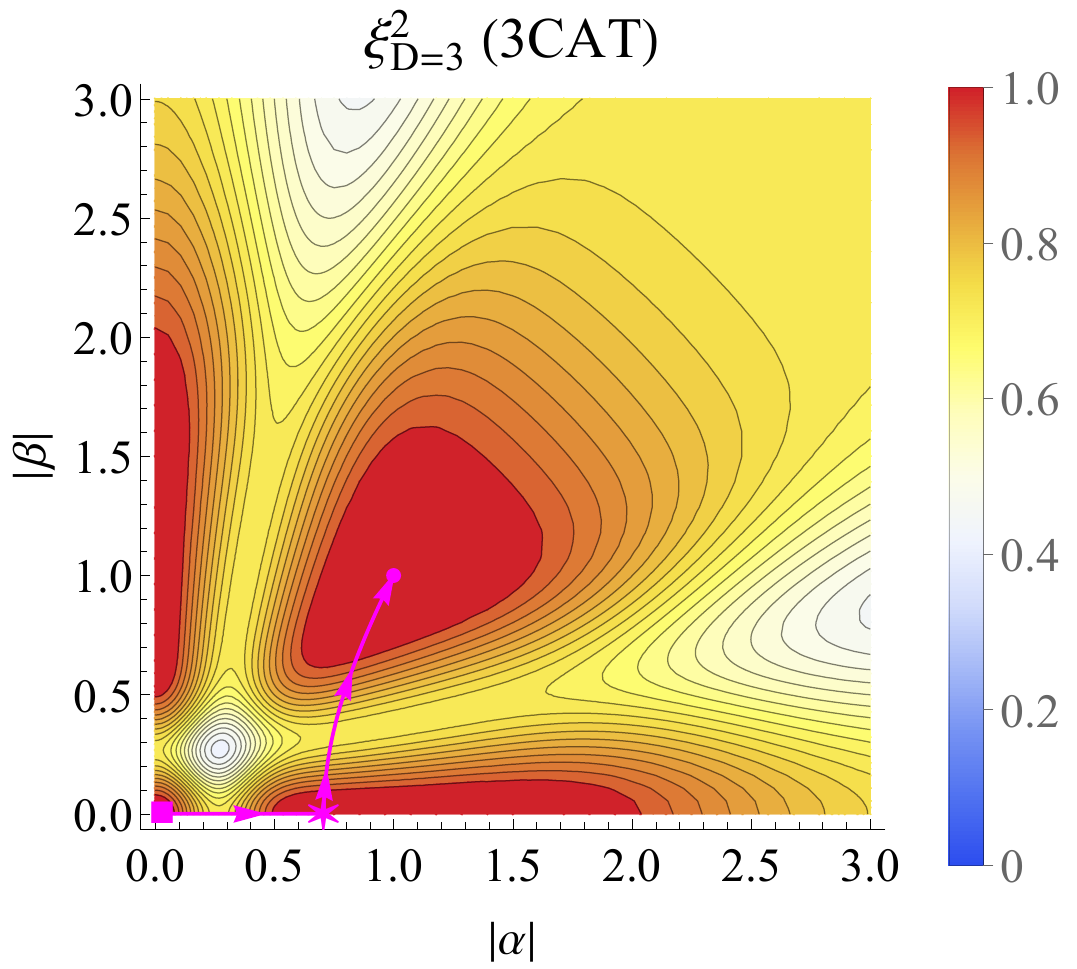}
\end{center}
\caption{Contour plots of the squeezing parameter $\xi^2_{D=3}$ of a $\rmu(3)$ Schr\"odinger cat for $N=10$ atoms, as a function 
of the phase-space coordinates $\alpha, \beta$ (it just depends on moduli). The meaning of the magenta curve is the same as in previous Figures.}
\label{squeezingcontour}
\end{figure}

Note that $D$-spin squeezing parameters $\xi_{ij}$ are constructed in terms of $D$-spin quadratic EVs, as the two-quDit RDM \eqref{rho2} and its purity \eqref{pur2qudit} do. Therefore, the deep relation between pairwise 
entanglement and spin squeezing revealed by  Wang-Sanders in \cite{PhysRevA.68.012101-Wang} for symmetric multi-qubit systems is extensible to symmetric multi-quDits in the sense proposed here. In Figure \ref{squeezingcontour} we show a contour plot of 
the total $D$-spin squeezing parameter $\xi^2_D$ for the $\3cat$. As in previous figures, the magenta curve represents the trajectory in phase space of the stationary points \eqref{critalphabeta}  of the energy surface \eqref{enersym} 
of the three-level LMG Hamiltonian \eqref{hamU3} 
as a function of the control parameter $\lambda$. See later around Figure \ref{squeezingfig} in Section \ref{signaturesec} for further discussion.

\section{LMG model for three-level atoms and its quantum phase diagram} \label{LMGsec}

In this section we apply the previous mathematical machinery to the study and characterization of the phase diagram of  quantum critical $D$-level Lipkin-Meshkov-Glick atom models. 
The standard case of  $D=2$ level atoms has already been studied in the literature (see e.g.  \cite{Calixto_2017}). We shall restrict ourselves to $D=3$ level atoms for practical calculations, 
although the procedure can be easily extended to general $D$. In particular, we propose the following LMG-type Hamiltonian 
\begin{equation}
H=\frac{\epsilon}{N}(S_{33}-S_{11})-\frac{\lambda}{N(N-1)}\sum_{i\not=j=1}^3 S_{ij}^2,\label{hamU3}
\end{equation}
written in terms of collective $\rmu(3)$-spin operators $S_{ij}$. Hamiltonians of this kind have already been proposed in the literature \cite{Kus,KusLipkin,Meredith,Casati,Saraceno} [see also 
\cite{nuestroPRE} for the role of mixed symmetry sectors in QPTs of multi-quDit LMG systems]. We place levels symmetrically about $i=2$, with  
intensive energy splitting per particle $\epsilon/N$. For simplicity, we consider equal interactions, with coupling constant $\lambda$, for atoms in different levels, 
and vanishing interactions for atoms in the same level (i.e., we discard interactions of the form $S_{ij}S_{ji}$). Therefore, $H$ is invariant under parity transformations $\Pi_j$ in \eqref{parityop}, 
since the interaction term scatters pairs of particles conserving the parity of the population $n_j$ in each level $j=1,\dots,D$. Energy levels have good parity, the ground state being an even state. 
We divide the two-body interaction in \eqref{hamU3} by the number of atom pairs $N(N-1)$ to make 
$H$ an intensive quantity, since we are interested in the thermodynamic limit $N\to\infty$. We shall see that parity symmetry is spontaneously broken in this limit. 

As already pointed long ago by Gilmore and coworkers \cite{GilmorePhysRevA.6.2211,RevModPhys.62.867}, coherent states constitute in general a powerful tool for rigorously studying the
ground state and thermodynamic critical properties of some physical systems. The energy surface associated to a  Hamiltonian density $H$  
is defined in general as the coherent state expectation value of the Hamiltonian density in the thermodynamic limit. In our case, the energy surface acquires the following form
\be
E_{(\alpha,\beta)}(\epsilon,\lambda)= \lim_{N\to\infty}\langle \zb|H|\zb\rangle= \epsilon\frac{  \beta  \bar{\beta }-1}{
\alpha  \bar{\alpha }+\beta  \bar{\beta }+1}
-\lambda\frac{ \alpha ^2 \left(\bar{\beta }^2+1\right)+\left(\beta ^2+1\right) \bar{\alpha }^2+\bar{\beta }^2+\beta ^2
}{\left(\alpha  \bar{\alpha }+\beta  \bar{\beta }+1\right)^2},\label{enersym}
\ee
where we have used DSCS EVs of linear  \eqref{CSEV} and quadratic  \eqref{qEV} powers of $D$-spin operators $S_{ij}$ 
[actually, linear powers are enough due to the lack of quantum spin fluctuations in the thermodynamic limit \eqref{nofluct}], and we have used the parametrization 
$\zb=(1,\alpha,\beta)$, as in eq. \eqref{S3C}, for $\rmu(3)$ SCSs. Note that this energy surface is invariant under $\alpha\to-\alpha$ and  $\beta\to-\beta$, which is a consequence of 
the inherent parity symmetry of the  Hamiltonian \eqref{hamU3}. 
\begin{figure}[h]
\begin{center}
\includegraphics[width=4cm]{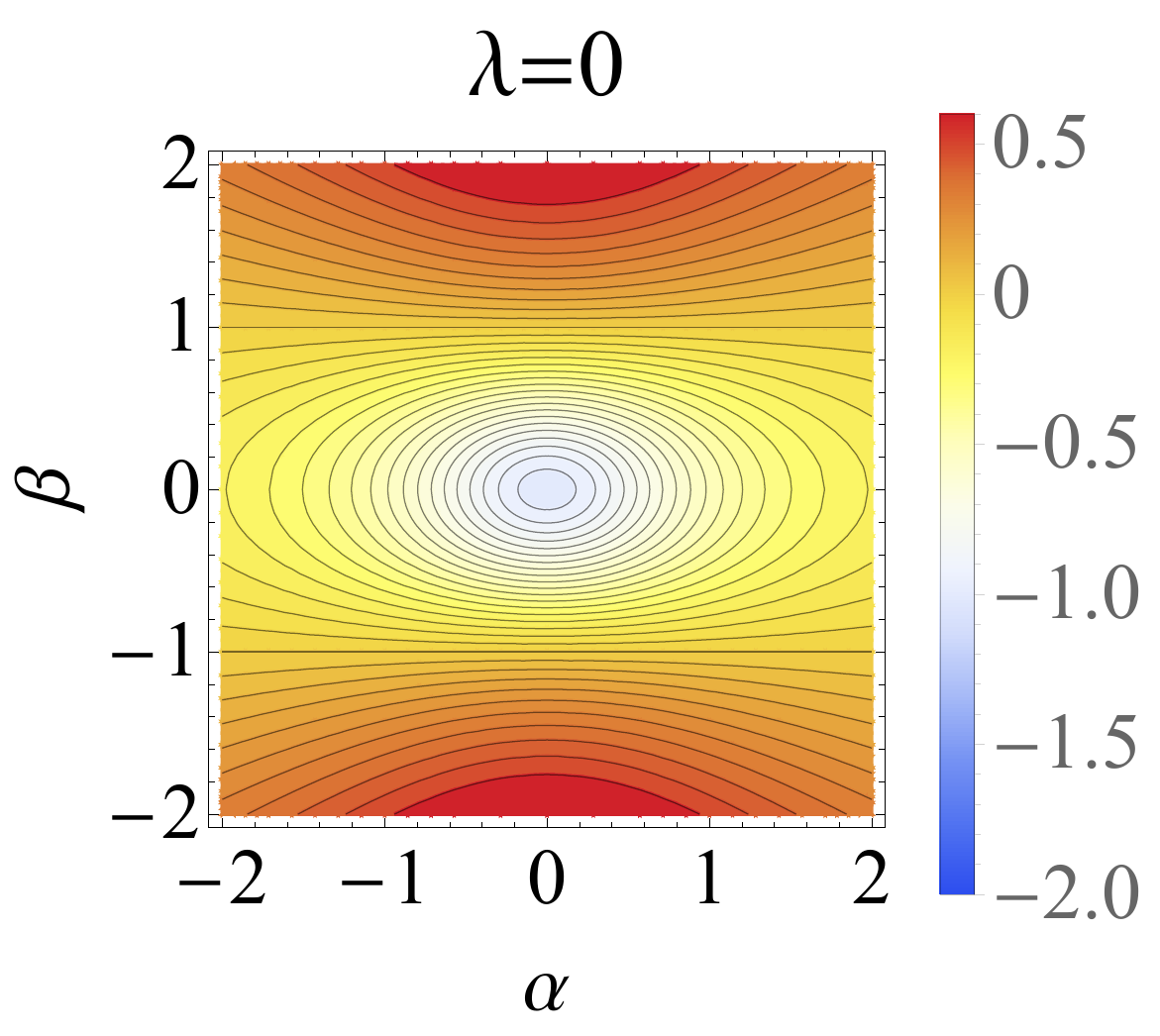}
\includegraphics[width=4cm]{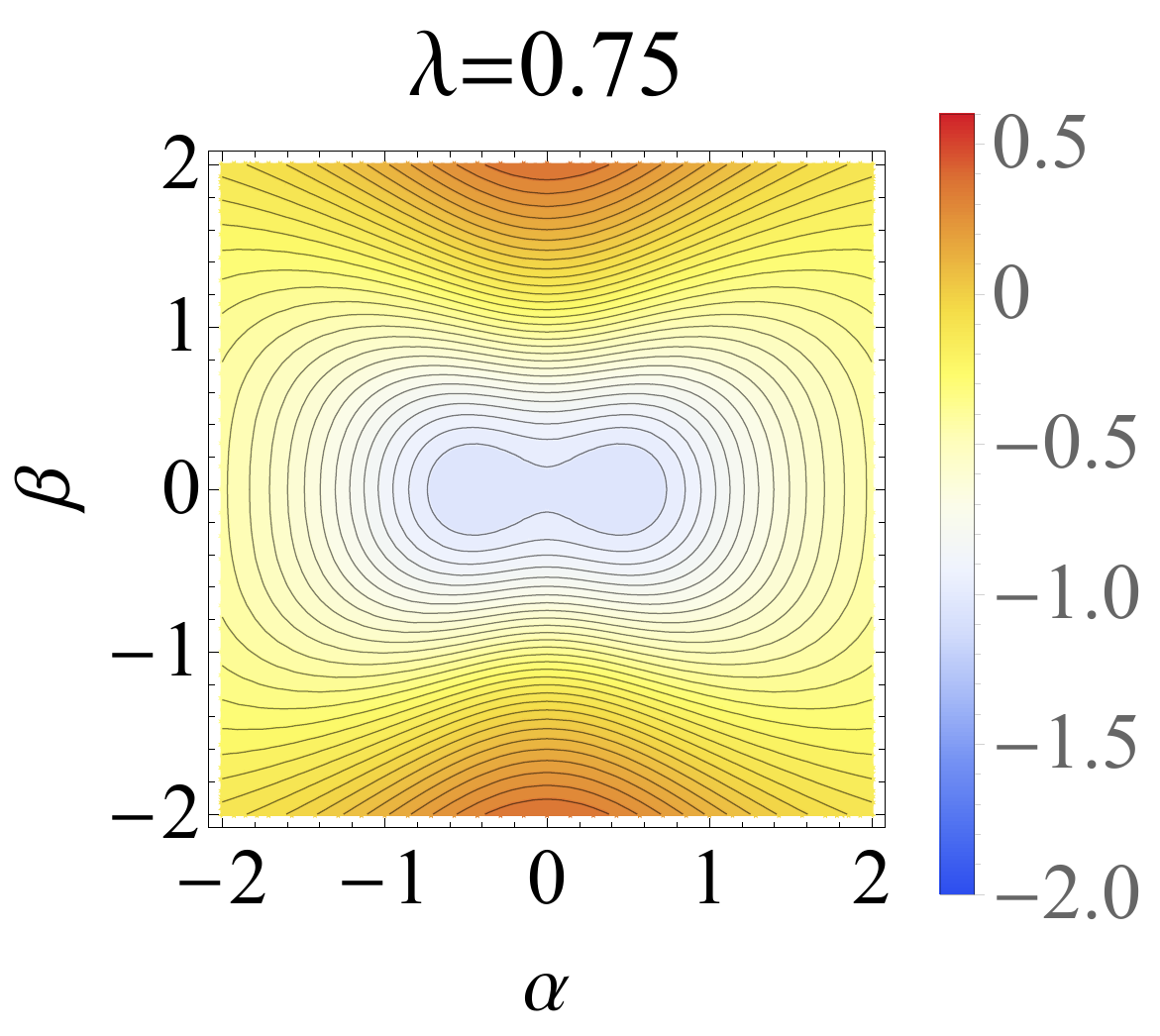}
\includegraphics[width=4cm]{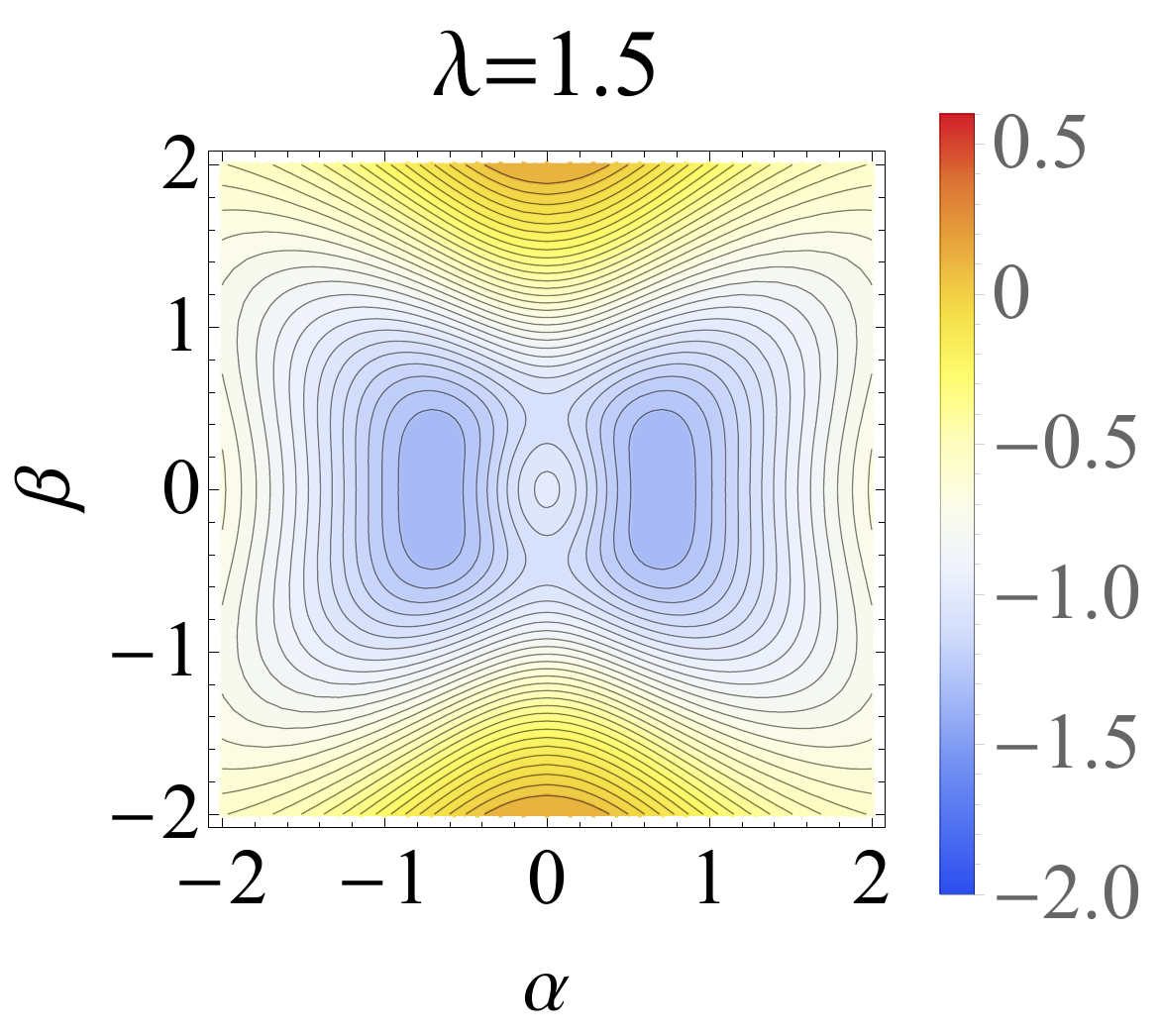}
\includegraphics[width=4cm]{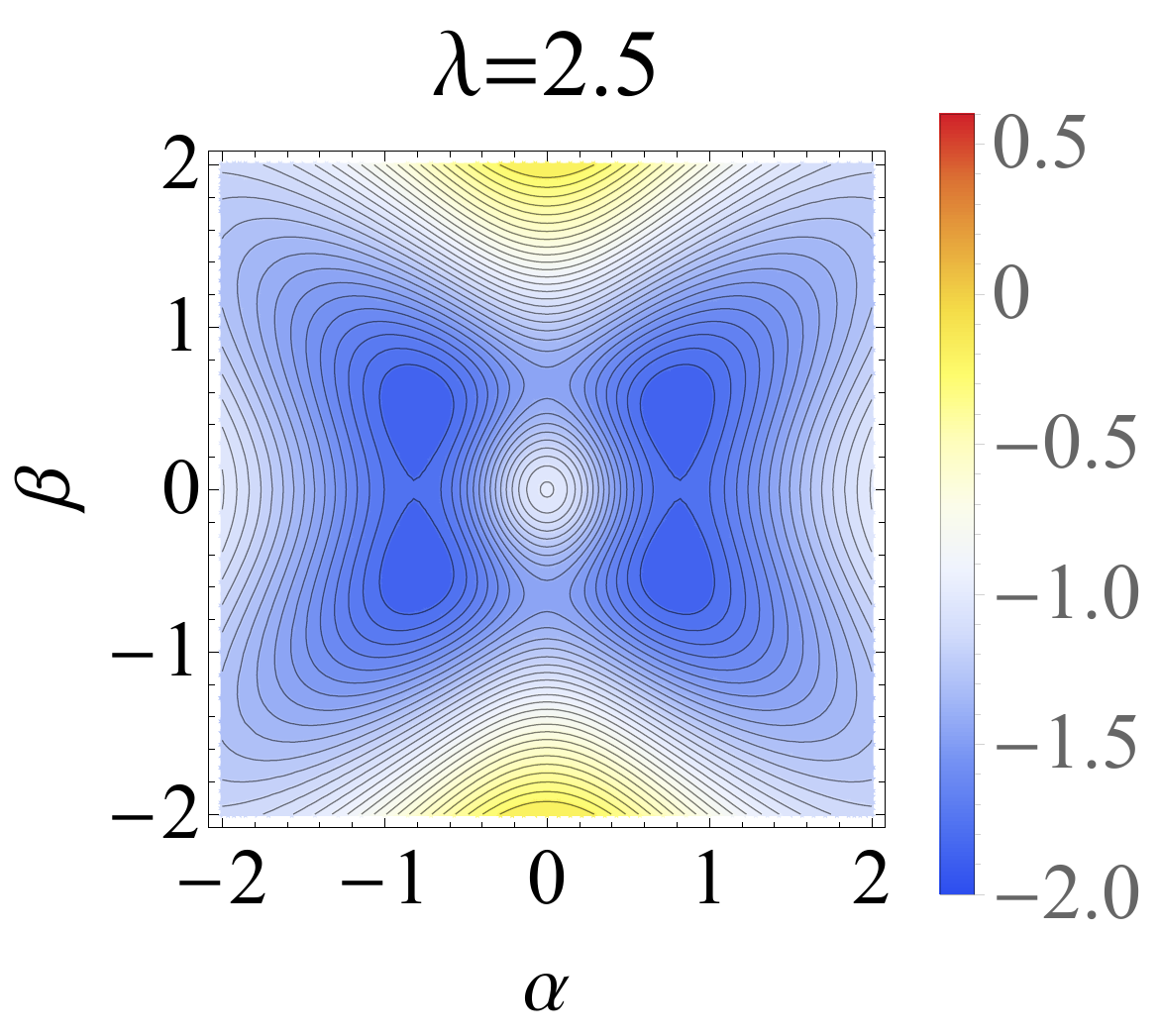}
\end{center}
\caption{Contour plot of the energy surface \eqref{enersym}  for real $\alpha$ and $\beta$, in the vicinity of the critical 
points $\lambda=1/2$ and $\lambda=3/2$ (in $\epsilon$ units). Degenerate minima are perceived in phases II 
$(\frac{1}{2}\leq \lambda \leq \frac{3}{2})$ and III $(\lambda \geq \frac{3}{2})$.}
\label{fig3}
\end{figure}
The minimum energy 
\be E_0(\epsilon,\lambda)=\mathrm{min}_{\alpha,\beta\in \mathbb{C}}E_{(\alpha,\beta)}(\epsilon,\lambda)\label{minimieq}\ee
is attained at the stationary (real) phase-space values $\alpha_0^\pm=\pm \alpha_0$ and $\beta_0^\pm=\pm\beta_0$ with
\bea
\alpha_0(\epsilon,\lambda)&=&\left\{\begin{array}{lll}
 0, && 0\leq \lambda \leq \frac{\epsilon }{2}, \\
 \sqrt{\frac{2\lambda- \epsilon }{2 \lambda +\epsilon }}, && \frac{\epsilon }{2}\leq \lambda \leq \frac{3 \epsilon }{2}, \\
 \sqrt{\frac{2\lambda }{2 \lambda +3 \epsilon }}, && \lambda \geq \frac{3 \epsilon }{2},
\end{array}\right.\nonumber\\
\beta_0(\epsilon,\lambda)&=&\left\{\begin{array}{lll}
 0, & & 0\leq \lambda \leq  \frac{3 \epsilon }{2}, \\
 \sqrt{\frac{2 \lambda -3 \epsilon}{2 \lambda +3 \epsilon }}, & & \lambda \geq \frac{3 \epsilon }{2}. \end{array}\right. \label{critalphabeta}
\eea
In Figures \ref{puritylevel3CAT}, \ref{purityOne3CAT}, \ref{purityTwo3CAT} and \ref{squeezingcontour} we plotted (in magenta color) the stationary-point curve 
$(\alpha_0(\lambda),\beta_0(\lambda))$ on top of level, one- and two-qutrit entanglement entropies, and squeezing parameter, noting that $(\alpha_0(\lambda),\beta_0(\lambda))\to (1,1)$ for high $\lambda\to\infty$ interactions. 
We will come to this later in Section \ref{signaturesec}. Inserting \eqref{critalphabeta} into  \eqref{enersym} gives the ground state energy density at the thermodynamic limit 
\be
E_0(\epsilon,\lambda)=\left\{\begin{array}{lllr}
 -\epsilon,  && 0\leq \lambda \leq \frac{\epsilon }{2}, & \mathrm{(I)}\\
 -\frac{(2 \lambda +\epsilon )^2}{8 \lambda }, && \frac{\epsilon }{2}\leq \lambda \leq \frac{3 \epsilon }{2}, &  \mathrm{(II)} \\
  -\frac{4\lambda^2+3\epsilon ^2}{6 \lambda }, & &\lambda \geq \frac{3 \epsilon }{2}. &  \mathrm{(III)}\end{array}\right.\label{energysym}
\ee
Here we clearly distinguish three different phases: I, II and III, and two second-order QPTs at $\lambda^{(0)}_{\mathrm{I}\leftrightarrow\mathrm{II}}=\epsilon/2$ and 
$\lambda^{(0)}_{\mathrm{II}\leftrightarrow\mathrm{III}}=3\epsilon/2$, respectively, where $\frac{\partial^2E_0(\epsilon,\lambda)}{\partial\lambda^2}$ are discontinuous. In the stationary (magenta) curve $(\alpha_0(\lambda),\beta_0(\lambda))$, the phase I corresponds to the origin $(\alpha_0,\beta_0)=(0,0)$ (squared point), phase II corresponds to the horizontal part $\beta_0=0$ up to the star point, and phase III corresponds to $\beta_0\neq 0$. 

Note that the ground state is fourfold degenerated in the thermodynamic limit since the four $\rmu(3)$ SCSs $|\zb_0^{\pm\pm}\ra=|1,\pm\alpha_0,\pm\beta_0\ra$ have the same energy density $E_0$. These four $\rmu(3)$ SCSs 
are related by parity transformations $\Pi_j$ in \eqref{parityop} and, therefore, parity symmetry is spontaneously broken in the thermodynamic limit. In order to have good variational states for finite $N$,  
to compare with numerical calculations, we have two possibilities: 1) either we use the 3CAT \eqref{S3C} as an ansatz for the ground state, minimizing $\langle\3cat|H|\3cat\rangle$, or 2) we restore 
the parity symmetry of the $\rmu(3)$ SCS $|1,\alpha_0,\beta_0\ra$ for finite $N$ by projecting on the even parity sector. Although the first possibility offers a more accurate variational approximation to the ground state, 
it entails a more tedious numerical minimization than the one already obtained in \eqref{minimieq} for $N\to\infty$. Therefore, we shall use the second  possibility which, despite being less accurate, it is 
straightforward and good enough for our purposes. That is, we shall use the 3CAT \eqref{S3C}, evaluated at $\alpha=\alpha_0$ and $\beta=\beta_0$ and conveniently normalized \eqref{S3CN}, as a variational approximation 
$|\3cat_0\ra$ to the numerical (exact) ground state $|\psi_0\ra$ for finite $N$.

\section{Entanglement and squeezing as signatures of  QPTs}\label{signaturesec}

The objective in this Section is to use level and particle entanglement and squeezing measures as signatures of QPTs in these LMG models, playing the role of order parameters that characterize the different phases or 
markers of the corresponding critical points. We restrict ourselves to linear entropy  which, as already shown, gives qualitative information similar to von Neumann entropy for this study, 
with the advantage that it requires less computational resources. As already commented, Refs. \cite{JPhysA.36.12255,PhysRevA.67.022110} contain more general information about the relation between both entropies. 
Linear entropies of one- and two-qutrit RDMs turn also to provide similar qualitative information, although pairwise 
entanglement shows a more direct relation to spin squeezing.

\begin{figure}[h]
\begin{center}
\includegraphics[width=12cm]{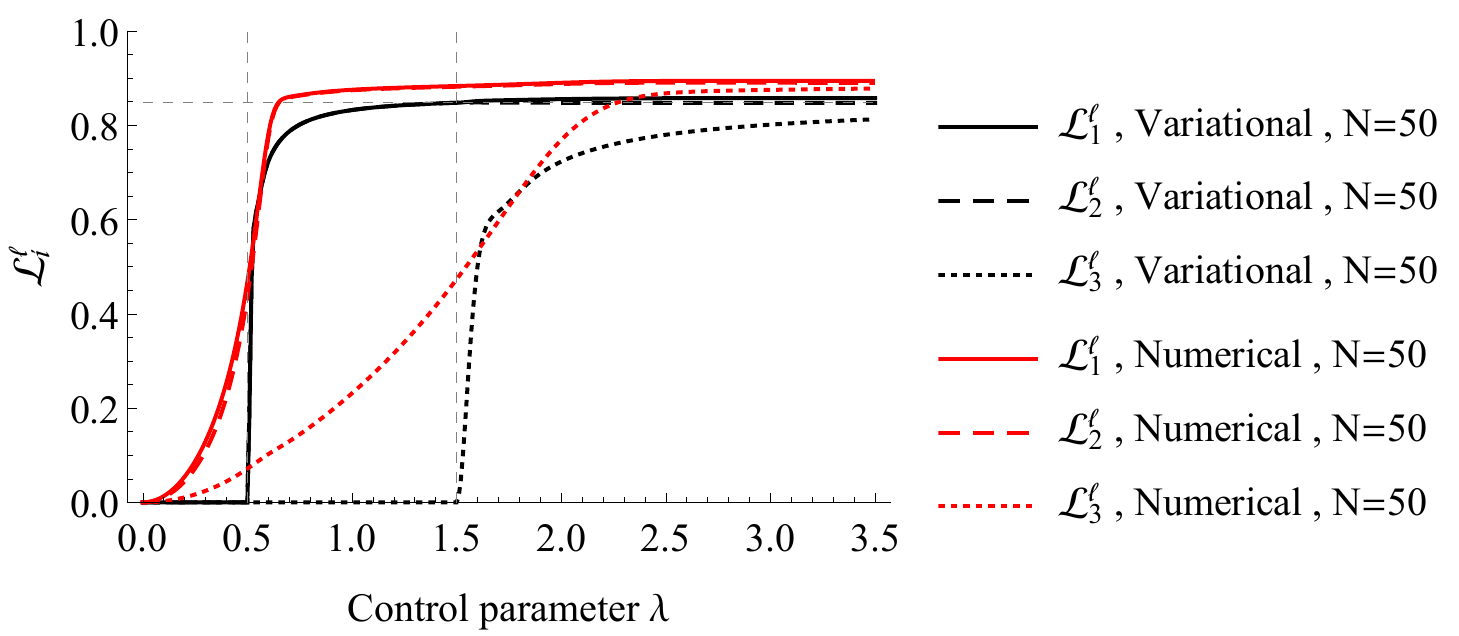}
\end{center}
\caption{Level entanglement linear entropies $\mathcal{L}_i^\ell$ (for levels $i=1,2$ and 3) of the ground state of the  three-level atom LMG model Hamiltonian \eqref{hamU3}, for $N=50$ atoms,   
as a function of the control parameter $\lambda$ (in $\epsilon$ units). 
Critical points, at which a QPT takes place,  are 
marked with vertical grid lines, whereas the horizontal grid line labels the  asymptotic value $\mathcal{L}_i^\ell\to 1-2/\sqrt{\pi N}\simeq 0.84$ of the entropies.  
We compare exact results, obtained from numerical diagonalization of the Hamiltonian, 
with variational (analytical)  results obtained from a parity symmetry restoration 
(in terms of Schr\"odinger cats) of mean field results.}
\label{puritylevelLMG}
\end{figure}

We have numerically diagonalized the Hamiltonian \eqref{hamU3} for $N=50$ 3-level atoms, and several values of $\lambda$ (in $\epsilon$ units), and we have calculated level, one- and two-qutrit entanglement 
linear entropies for the ground state $|\psi_0\ra=\sum_{\vec{n}}'{}\,c_{\vec{n}}|\vec{n}\ra$, plugging the coefficients $c_{\vec{n}}$ into (\ref{isoEV},\ref{purylevel},\ref{pur2qudit}). We have also 
calculated level and atom entanglement linear entropies for the variational approximation $|\3cat_0\ra$ to the ground state $|\psi_0\ra$ discussed in the previous Section for $N=50$. 
In Figure \ref{puritylevelLMG} we compare 
numerical with variational ground state entanglement measures between levels $i=1,2,3$. According to Figure \ref{puritylevelLMG}, we see that, in phase I, $0\leq \lambda \leq 1/2$,  
 variational results indicate that there is no entanglement between levels, whereas numerical 
results show a small (but non-zero) entanglement for $N=50$.  In phase II, $1/2\leq \lambda \leq 3/2$, levels $i=1$ and $i=2$ get entangled, but level $i=3$ remains almost disconnected.  
In phase III, $\lambda \geq 3/2$, level $i=3$ gets entangled too. Interlevel entanglement grows with $\lambda$ attaining the maximum value of 0.84 at the limiting 
point $(\alpha_0(\infty),\beta_0(\infty))=(1,1)$ for $N=50$. This behavior of the interlevel entropy for the $\3cat$ variational state can be also appreciated by looking at the stationary (magenta) curve in Figure \ref{puritylevel3CAT} with relation to the isentropic 
curves.

\begin{figure}[h]
\begin{center}
\includegraphics[width=12cm]{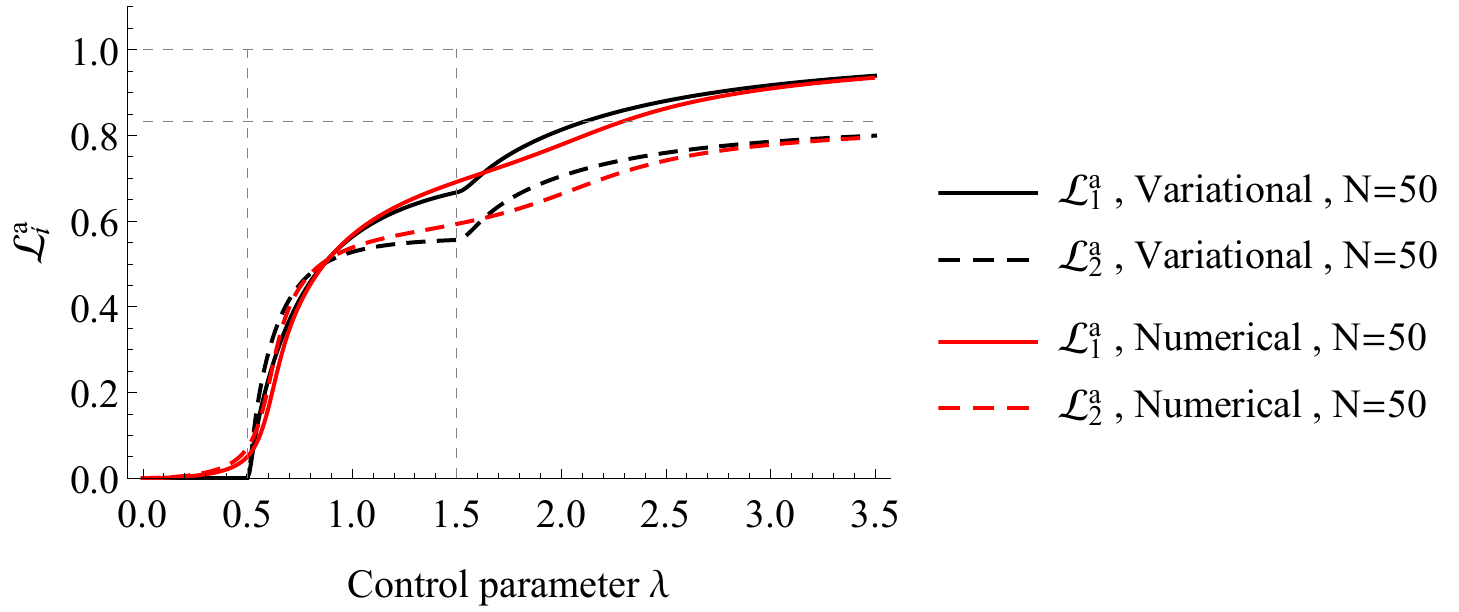}
\end{center}
\caption{One-qutrit $\mathcal{L}_1^\mathrm{a}$ and two-qutrit $\mathcal{L}_2^\mathrm{a}$ entanglement linear entropies of the ground state of the  three-level atom LMG model Hamiltonian \eqref{hamU3}   
as a function of the control parameter $\lambda$ (in $\epsilon$ units). 
Critical points, at which a QPT takes place,  are 
marked with vertical grid lines, whereas horizontal grid lines label the  asymptotic values $\mathcal{L}_1^\mathrm{a}\to 1$ and $\mathcal{L}_2^\mathrm{a}\to 5/6$ of the entropies.  
We compare numerical with variational results for $N=50$ atoms. }
\label{purityatomLMG}
\end{figure}

Concerning atom entanglement,  Figure \ref{purityatomLMG} shows a better agreement between variational and numerical results, showing 
a rise of entanglement when the coupling strength $\lambda$ grows across the three phases, attaining values close to the large $N$ maximum values $\mathcal{L}_1^\mathrm{a}=1$ and $\mathcal{L}_2^\mathrm{a}=5/6$
at the limiting point $(\alpha_0(\infty),\beta_0(\infty))\to(1,1)$. The entanglement growth  is more abrupt between phases I and II than between phases II and III. We see that both, level and atom 
entanglement measures capture  differences between the three phases, even for finite $N$, and therefore they can be considered as precursors of the corresponding QPT. The main features of the inter-atom entanglement entropy for the $\3cat$ variational state are also captured by the trajectory of the stationary (magenta) curve in Figures \ref{purityOne3CAT} and \ref{purityTwo3CAT} 
through the isentropic curves.

\begin{figure}[h]
\begin{center}
\includegraphics[width=12cm]{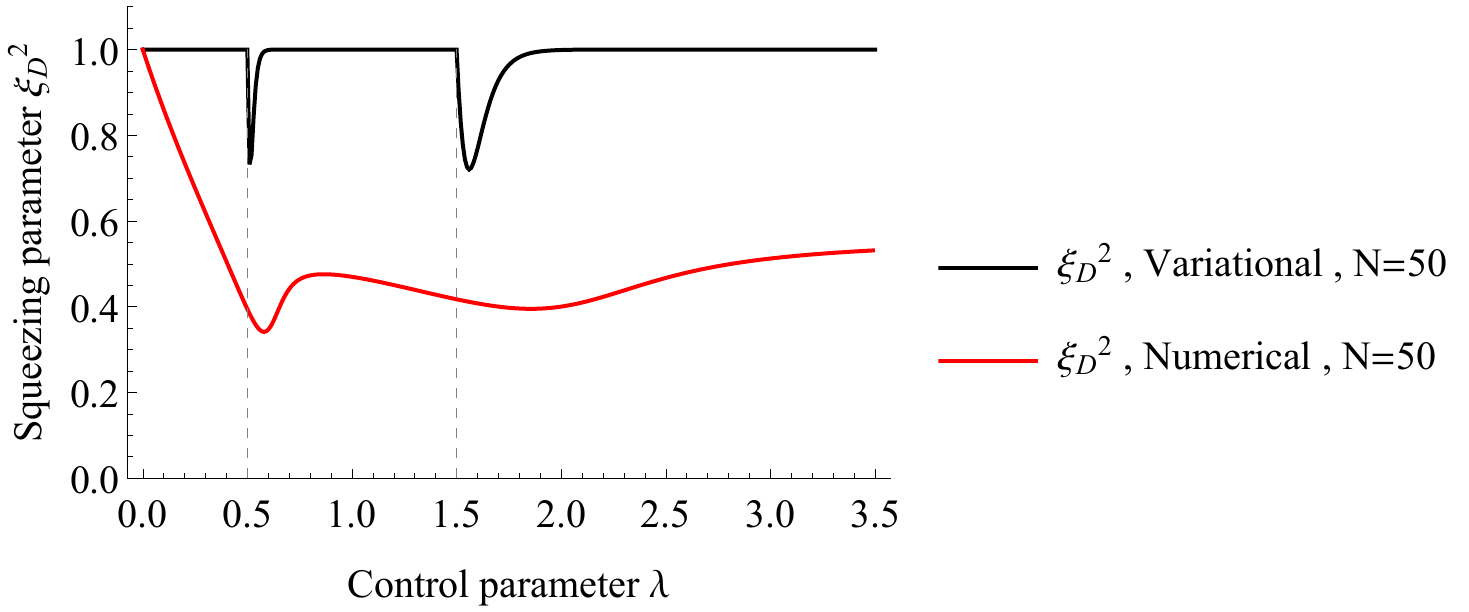}
\end{center}
\caption{$D$-spin total squeezing parameter $\xi^2_D$ \eqref{totalsqueezing}  of the ground state of the  three-level atom LMG model Hamiltonian \eqref{hamU3}  
as a function of the control parameter $\lambda$ (in $\epsilon$ units). Critical points, at which a QPT takes place,  are 
marked with vertical grid lines. We compare numerical with variational results for $N=50$ atoms.}
\label{squeezingfig}
\end{figure}

In Figure \ref{squeezingfig} we represent the $D=3$ spin total squeezing parameter $\xi^2_D$ \eqref{totalsqueezing}  of the variational and numerical ground states for $N=50$ atoms, 
as a function of the control parameter $\lambda$ (in $\epsilon$ units). The results reveal a clear growth of squeezing  at the critical points, the change being more abrupt 
at these points for the variational (parity adapted mean field)  than for the numerical ground state. Note that the variational ground state only shows squeezing ($\xi^2_D<1$) at the critical points, whereas the 
numerical ground state exhibits squeezing for any $\lambda\not=0$. Looking at the stationary (magenta) curve of Figure \ref{squeezingcontour} we appreciate that it 
practically lies in regions of no squeezing (in red color) except near the critical points, where squeezing suddenly increases (yellow color regions).

\section{Conclusions and outlook}\label{conclusec}
We have extended the concept of pairwise entanglement and spin squeezing for symmetric multi-qubits (namely, identical two-level atoms) to general symmetric multi-quDits (namely, identical $D$-level atoms). 
For it, we have firstly computed expectation values of $\rmu(D)$ spin operators $S_{ij}$ in general symmetric multi-quDit states like: $\rmu(D)$-spin coherent states, 
their adaptation to parity (Schr\"odinger $\dcat$ states), and an extension of $\noon$ states to $D$ levels ($\nodon$ states). The reduced density matrices to one- and two-quDits extracted at random 
from a symmetric multi-quDit state exhibit atom entanglement for $\dcat$ states, but not for $\rmu(D)$-spin coherent states. We have used entanglement to characterize quantum phase transitions 
of LMG $D$-level atom models (we have restricted to $D=3$ for simplicity), where $\dcat$ states (as an adaptation to parity of mean-field spin coherent states) turn out to be a reasonable good 
variational approximation to the exact (numerical) ground state. We have also proposed an extension of standard SU(2)-spin squeezing to SU$(D)$-spin operators, which recovers  $D=2$ as a particular case. 
We have evaluated SU$(3)$-spin squeezing of the ground state of the LMG 3-level atom model, as a function of the control parameter $\lambda$, and we have seen that squeezing grows in the neighborhood 
of critical points $\lambda_c$, therefore serving as a marker of the corresponding quantum phase transition. A deeper study and discussion of squeezing in these models requires a phase space approach 
in terms of a coherent (Bargmann) representation  of states, such as the Husimi and Wigner functions, and it will be the subject of future work.

\section*{Acknowledgments}
We thank the support of the Spanish MICINN  through the project PGC2018-097831-B-I00 and  Junta de Andaluc\'\i a through the projects SOMM17/6105/UGR, UHU-1262561 and FQM-381. 
JG also thanks MICINN for financial support from FIS2017-84440-C2-2-P. 
AM thanks the Spanish MIU for the FPU19/06376 predoctoral fellowship. We all thank Octavio Casta\~nos for his valuable collaboration in the early stages of this work.

\bibliography{/home/usuario/MEGA_Genfimat/bibliografia.bib}

\end{document}